# The i3+3 Design for Phase I Clinical Trials

Meizi Liu
Department of Public Health Sciences,
The University of Chicago
5841 South Maryland Ave MC2000 Chicago, IL 60637 – 1447

Sue-Jane Wang[1]
Center for Drug Evaluation and Research, US FDA
10903 New Hampshire Ave., HFD-700, WO 21, Room 3526 Silver Spring, MD 20993

Yuan Ji
Department of Public Health Sciences,
The University of Chicago
5841 South Maryland Ave MC2000 Chicago, IL 60637 - 1447 1

**Abstract**

**Purpose** The 3+3 design has been shown to be less likely to achieve the objectives of phase I dose-finding trials when compared with more advanced model-based designs. One major criticism of the 3+3 design is that it is based on simple rules, does not depend on statistical models for inference, and leads to unsafe and unreliable operating characteristics. On the other hand, being rule-based allows 3+3 to be easily understood and implemented in practice, making it the first choice among clinicians. Is it possible to have a rule-based design with great performance?

**Methods** We propose a new rule-based design called i3+3, where the letter "i" represents the word "interval". The i3+3 design is based on simple but more advanced rules that account for the variabilities in the observed data. We compare the operating characteristics for the proposed i3+3 design with other popular phase I designs by simulation.

**Results** The i3+3 design is far superior than the 3+3 design in trial safety and the ability to identify the true MTD. Compared with model-based phase I designs, i3+3 also demonstrates comparable performances. In other words, the i3+3 design possesses both the simplicity and transparency of the rule-based approaches, and the superior operating characteristics seen in model-based approaches. An online R Shiny tool (https://i3design.shinyapps.io/i3plus3/) is provided to illustrate the i3+3 design, although in practice it requires no software to design or conduct a dose-finding trial.

**Conclusion** The i3+3 design could be a practice-altering method for the clinical community.

# 1 INTRODUCTION

## 1.1. Background

Phase I dose-finding trials represent the first stage of testing a new drug or new therapy in humans and are crucial in clinical drug development as they provide dose recommendation for later phase clinical trials. The primary goal of phase I trials is to assess dose limiting toxicities (DLT) and find the maximum tolerated dose (MTD) while maintaining patient safety. Statistical designs for phase I dose-finding trials can be generally divided into two classes: rule-based methods and model-based methods. The most widely used 3+3 design (1) is an example of rule-based methods, in which dose escalation and de-escalation decisions are based on a set of prespecified rules. It is by far the most popular method in practice due to its simplicity.

Clinicians and drug developers only need to follow a set of rules (2) to design and conduct the dose-finding trials without needing to understand statistical models or run computer programs. The other class of designs uses parametric or nonparametric statistical models to describe the underlying mechanism of observed dose-toxicity relationship. Statistical inference based on the models leads to desirable decisions for dose allocation, typically optimized according to a certain criterion. The continuous reassessment method (CRM) by O' Quigley et al. (3) is an example of model-based designs. In practice, the CRM design often requires expertise in statistical modeling and the ability to execute computer programs for real-trial applications.

In addition to the classical model-based designs like CRM, there is an increasing interest in a type of model-based designs, called interval-based designs that demonstrate comparable operating characteristics to the classical model-based designs but with simpler implementation like the rule-based designs. Some notable examples include the mTPI (4) and mTPI-2 designs (5), the cumulative cohort design (CCD) (6) and the Bayesian optimal interval design (BOIN) (7). One important advantage of these designs is that the decisions of dose allocation can be pre-tabulated in advance, making the decision process transparent and intuitive to clinicians. However, they all rely on underlying models and inference that require expertise in statistics and mathematics.

### 1.2. A motivating example

The proposed i3+3 design is motivated by a debatable decision based on the mTPI design (5). We first provide a brief review of the mTPI design to set the stage for the upcoming discussion. The mTPI design requires investigators to provide two input values, the target probability of toxicity $p_T$ and the equivalence interval (EI), defined mathematically as $[p_T - \varepsilon_1, p_T + \varepsilon_2]$. The target probability of toxicity $p_T$ specifies the highest toxicity rate of the MTD (e.g., $p_T$=0.3) and the EI provides a range around the target probability so that doses with toxicity probabilities inside the EI will also be considered as the MTD. For example, when $p_T$=0.3 and EI is [0.25, 0.35], the target toxicity rate of the MTD is 0.3, but a dose with toxicity rate between 0.25 and 0.35 will also be considered as the MTD. The use of EI allows some variabilities in the statistical inference and flexibility for real-world clinical trials.

Statistically, mTPI uses an optimal decision rule for assigning patients (4). However, some decisions of mTPI, even though optimal and accounting for the data variability, may be considered risky in clinical practice. For example, suppose in a dose-finding trial, the target toxicity probability is $p_T = 0.3$ and the EI is [0.25, 0.35]. This means that the MTD should have a toxicity probability near 0.3 but not below 0.25 or above 0.35. Suppose 6 patients have been enrolled at a dose and 3 of them experience DLTs at a dose currently being used for treatment. Based on the observed data, a trial design needs to provide a decision to select the dose level at which the next cohort of patients will be assigned. The decision of the mTPI design is "S", to stay and enroll more patients at the current dose. The decision "S" can be deemed too aggressive since the observed toxicity rate is 3/6= 0.5, which is higher than the target rate $p_T = 0.3$. In practice, the decision "D", to de-escalate the dose level is considered to be safer and more desirable. Through personal communication, this type of argument has been raised by IRB review committees and regulatory agencies. However, a question one may ask is that why a mathematically optimal decision rule in mTPI produces the decision "S" that is perceived too risky in practice?

To address this question, we begin by noting that there is no consensus on what decisions are acceptable in real-world trials. Oftentimes, whether a decision is acceptable depends on the review committee's experiences, preference, and common sense. For example, if 3 out of 3 patients experience DLTs at a dose, the only acceptable decision is "D", to de-escalate the dose level regardless of what statistical design is used. However, it is less obvious which decisions are acceptable when 3 out of 6 patients have DLTs in a trial with a target toxicity rate $p_T$=0.3 and EI = [0.25, 0.35]. So why is decision "S" considered not acceptable by review committees here? The most obvious answer would be that the observed toxicity rate 3/6=0.5 is higher than the target rate 0.3. An observed toxicity rate higher than the MTD target rate raises safety concern of the dose, and therefore the decision "S", which allows more patients to be treated at the same dose, may be considered too risky in practice. However, this may not be the only reason. Let us consider another example. The same decision "S" is used by the 3+3 design if 1 out of 3 patients experiences DLT at a dose. The 3+3 design defines the MTD as the highest dose with no more than 1 out of 6 patients experiencing DLT. In other words, the target toxicity rate of the 3+3 design is about 1/6. When 1 out of 3 patients experiences DLT, the observed toxicity rate is 1/3, which is higher than 1/6, and yet the decision "S" under the 3+3 design is widely accepted in practice. So why is decision "S" acceptable when 1 out of 3 patients experiences DLT with a target rate 1/6 but not acceptable when 3 out of 6 patients have DLTs with a target rate 0.3?

The answer lies in the difference of the sample sizes (and associated data variabilities) of the observed data, 6 patients versus 3 patients, between the two examples. Consider the 3+3 case where "S" is acceptable when 1 out of 3 patients has DLT and the target rate is 1/6. Even though the observed toxicity rate 1/3 is twice higher than the target rate 1/6, there is only 3 patients worth of information, which does not sufficiently distinguish 1/3 from 1/6. For example, with 1 fewer DLT, the observed rate would be 0/3, way below 1/6. In contrast, in the second case where 3 out of 6 patients have DLT and the target rate is 0.3, not only is the empirical rate 3/6 higher than the target rate, but even if 1 fewer patient were to have DLT, the observed rate would be 2/6 which would still be higher than 0.3. Therefore, the observed rate 3/6 in the second case is more informative than the observed rate 1/3 in the first case, despite both observed rates being higher than the target rates. As a result, the decision "S" is considered not acceptable in the case of 6 patients but acceptable in the case of 3 patients.

Motivated by this reasoning, we propose a new rule-based design, the i3+3 design in which the letter "i" stands for "interval". The main innovation of i3+3 is to add an additional criterion into the 3+3 like rules so that data variability is accounted for. The new criterion considers the gap between two possible data points and its relative magnitude to the EI.

## 2 The PROPOSED i3+3 DESIGN

### 2.1. Dose-finding algorithm

Suppose a set of $d = 1, \ldots D$ ascending doses is tested in a trial. Assume the target toxicity rate $p_T$ for the MTD (e.g., $p_T = 0.3$) and the EI (e.g., EI = [0.25, 0.35]), have been specified and fixed for the trial. Suppose dose $d$ is the dose currently used to treat patients, $n$ is the number of patients who have been treated at dose $d$, and $x$ is the number of patients who have experienced DLTs at dose $d$. Assume patients are enrolled and assigned to doses in cohorts. That is, the next cohort of patients may not be enrolled until toxicity outcomes from the previous cohorts of enrolled patients have been fully observed. Given the observed data, the i3+3 design identifies an appropriate dose for the next cohort of patients. We propose a set of rules in Table 1 that fulfills the task. Specifically, these rules compare two quantities, $\frac{x}{n}$ and $\frac{x-1}{n}$, to the EI, and depending on the comparison outcome, decide the next dose level for enrolling patients.

In words, the i3+3 design can be summarized as follows: at the current dose $d$, calculate two quantities $\frac{x}{n}$ and $\frac{x-1}{n}$.

- If $\frac{x}{n}$ is below the EI, escalate ("E") and enroll patients at the next higher dose ($\underline{d+1}$);
    - else, if $\frac{x}{n}$ is inside the EI, stay ("S") and continue to enroll patients at the current dose $\underline{d}$;
        - else, if $\frac{x}{n}$ is above the EI, there are two scenarios:
            1) if $\frac{x-1}{n}$ is below the EI, stay ("S") and continue to enroll patients at the current dose $\underline{d}$,
            2) else, de-escalate ("D") and enroll patients at the next lower dose ($\underline{d-1}$).

When $d$ is the highest dose or the lowest dose, the above rules need to be modified as special cases.

1. If the current dose is the highest dose, and $\frac{x}{n}$ is below the EI, stay ("S") instead of escalating ("E") and continue to enroll patients at the current dose. This is because there is no dose to escalate to.

2. If the current dose is the lowest dose, and $\frac{x}{n}$ is above the EI, stay ("S") instead of potentially de-escalating ("D") and continue to enroll patients at the current dose. This is because there is no dose to de-escalate to.

Following the mTPI and mTPI-2 designs, we add two safety rules as ethical constraints to avoid excessive toxicity:

- Safety rule 1 (<u>early termination</u>): Suppose that dose 1 has been used to treat patients. If $Pr(p_1 > p_T | x_1, n_1) > \xi$ for $\xi = 0.95$, terminate the trial due to excessive toxicity.
- Safety rule 2 (<u>dose exclusion</u>): Suppose that the decision is to escalate from dose $d$ to $(d+1)$. If $Pr(p_{d+1} > p_T | x_{d+1}, n_{d+1}) > \xi$ for $\xi = 0.95$, then treat the next cohort of patients at dose $d$ instead of $(d+1)$, and dose $(d+1)$ and higher doses will be removed from the trial. These doses will be marked as "DU" to indicate that they are Unacceptable to use in the trial anymore.

Under the i3+3 design, a trial is terminated either when a prespecified sample size is reached or according to Safety rule 1.

## 2.2. Examples and Software

In Table 2, we provide two cases to illustrate the application of the i3+3 design. In case 1, $n = 6$ patients are treated at dose $d$, $p_T = 0.3$ and the EI is $[0.25, 0.35]$. In case 2, $n = 3$, $p_T = 0.17$ and the EI is $[0.12, 0.22]$. We also include the decisions of mTPI and mTPI-2, two interval-based designs for comparison. In either case, the three designs agree in all decisions except for one. In case 1, the i3+3 decision for $x = 3$ is "D", to de-escalate to dose $(d - 1)$. With the same setting, the mTPI decision is "S", to continue treating at dose $d$. This decision is considered too aggressive in practice according to our previous discussion. In case 2, the i3+3 decision for $x = 1$ is "S", to continue treating at dose $d$ which coincides with the 3+3 design. However, the mTPI-2 decision is "D", to de-escalate to dose $(d - 1)$, which may not be preferred by physicians who side with the 3+3 design. In both cases, the i3+3 design seems to make the "humanly desirable" decisions as opposed to the model-based designs like mTPI and mTPI-2.

In addition, based on the proposed i3+3 algorithm in Table 1, we can easily generate decisions for any number of $n$ patients, not just restricted to 3 or 6. We provide an R Shiny tool freely available at https://i3design.shinyapps.io/i3plus3/ that generates decision tables based on the i3+3 design for any $p_T$ and EI values. See Figure 1 for an example for a trial with $p_T = 0.3$ and EI =[0.25, 0.35]. In the decision table, we present decisions in terms of three letters, "D", "S", and "E", which correspond treating future patients at dose level $(d-1)$, $d$, and $(d+1)$, respectively. The letter "U" represents the situations where Safety rule 2 is invoked. This table makes the i3+3 decisions transparent to the investigators prior to trial start and can be used to guide all the dose assignment decisions throughout the trial. Alternatively, investigators can memorize and follow the simple rules in Table 1 to guide dose assignments in the trial, which only requires comparing $\frac{x}{n}$ and $\frac{x-1}{n}$ to the EI. The i3+3 is sufficiently simple and intuitive for this alternative choice.

In addition, the R Shiny tool conducts simulation based on the i3+3 design and allows investigators to examine the operating characteristics of the design, which will be discussed next.

## 2.3. Estimation of MTD

When a trial is terminated, the MTD selection under the i3+3 design follows the same procedure as in the mTPI and mTPI-2 designs. Specifically, let $p_T$ denote the target toxicity rate, and $\epsilon_1, \epsilon_2$ are small positive constants such that EI= $[p_T - \epsilon_1, p_T + \epsilon_2]$. For example, $\epsilon_1 = \epsilon_2 = 0.05$. Let $p_d$ denote the probability of toxicity at dose $d$. Assume $n_d$ patients have been treated at dose $d$, and $x_d$ of them experienced DLTs. Therefore, we assume $x_d|p_d \sim Bin(n_d, p_d)$, follows a binomial distribution. In addition, assume the prior distribution $p_d \sim Beta(0.005, 0.005)$ as in the mTPI and mTPI-2 designs. These parameter values of 0.005 are chosen so that the beta priors have little impact on the posterior distribution, which is given by $p_d|(x_d, n_d) \sim Beta(x_d + 0.005, n_d - x_d + 0.005)$. Therefore, the posterior mean toxicity probability at dose $d$ is given by

$$\hat{p}_d = \frac{x_d + 0.005}{n_d + 0.01}$$

To impose the monotonicity of the dose-toxicity relationship, isotonic regression is applied to the posterior means $\hat{p}_d$ using the pool adjacent violators algorithm (8). This algorithm replaces adjacent estimates that violate the monotonicity assumption with their weighted average, where the weights are the posterior variance at each dose level. Let $\tilde{p}_d$ be the isotonic-transformed posterior means for all the doses. The estimated MTD is the dose with the smallest difference

$$d^* = \underset{d \in \{1,2,\ldots D\}}{\operatorname{argmin}} |\tilde{p}_d - p_T|$$

among all the doses $d$ for which $n_d > 0$ and $\tilde{p}_d \leq (p_T + \epsilon_2)$.

If more than one dose of $d^*$ exists, it means there are ties for which the value $\tilde{p}_{d^*}$ is the same. Then we choose MTD according to the following simple rules:

a) If $\tilde{p}_{d^*} > p_T$, the suggested MTD is the lowest dose in $d^*$.

b) If $\tilde{p}_{d^*} \leq p_T$, the suggested MTD is the highest dose in $d^*$.

## 3 OPERATING CHARACTERISTICS OF THE i3+3 DESIGN

### 3.1 Simulation Setup

Via repeated simulation trials, we assess and compare the operating characteristics of the i3+3 design. For comparison, we also include the 3+3, mTPI, mTPI2, BOIN, CRM and the Bayesian Logistic Regression Model (BLRM) (9) designs in the simulation. Simulated trials are generated based on scenarios that specify the true toxicity rate of each dose. We slightly modified the three sets of 14 scenarios in (10) that assume six doses for a trial and three different target $p_T$ values of 0.1, 0.17, and 0.3, respectively. Scenarios for $p_T = 0.17$ are generated by subtracting 0.03 from the original scenarios for $p_T = 0.2$ in (10). See Appendix A for details. This gives a total of 42 scenarios. A sample size of 30 patients is used for each trial with a cohort size of 3. For the mTPI and mTPI-2 designs, the EI is $[p_T - 0.05, p_T + 0.05]$. For the BOIN design (8), we set $(p_T - \lambda_e) = 0.05$ and $(\lambda_d - p_T) = 0.05$ so that the actual intervals for decision making among BOIN, mTPI and mTPI-2 are identical. For the BLRM design, the EWOC parameter is set at 0.5. A detailed description of the CRM and BLRM used in the simulation study is provided in Appendix B. Under each simulation scenario, we generate 1,000 simulated trials.

### 3.2 Simulation Results

We characterize the true MTDs as any doses with toxicity probabilities inside the EI. If no doses satisfy the criterion, the true MTD is the highest dose among those with true toxicity probabilities $p_d < p_T$. If no MTD could be identified based on the two criteria, then the particular scenario does not have a true MTD, which means the decision of selecting any dose as the MTD would be wrong.

To assess the performance of a dose-finding design, we consider the following criteria:
1. Reliability or the percentage of correct selection (PCS), defined as the percentage of the trials that the MTD is correctly selected. When all the dose levels are above the MTD, PCS is defined as the percentage of early termination of trials.
2. Safety, defined as the average percentage of the patients treated at or below the true

MTDs across the simulated trials.

A desirable dose-finding design should demonstrate good balance between patient safety and ability to identify the true MTD. Additional results such as the percentage of trials concluding a dose above the true MTD or percentage of stopping a trial before reaching maximum sample size are given in Appendix C.

To summarize the simulation results, we take pair-wise differences between the i3+3 design and another design under comparison in their safety and reliability values for the same scenario, and display the results through the boxplots in Figure 2. A positive value means the design under comparison is safer or more reliable than the i3+3 design. As can be seen, BLRM with EWOC=0.5 is a safer but ultra-conservative design than i3+3. This has also been observed in the recent work (11). In general, the i3+3 design demonstrates comparable safety performance with mTPI, mTPI-2, BOIN and CRM, and the 3+3 design shows the worst safety among all the designs under comparison. In terms of reliability, the i3+3 design performs similarly with the mTPI, mTPI-2, BOIN and CRM designs, while the BLRM and 3+3 designs are the least reliable in the identification of the true MTDs. Overall, the i3+3 design shows comparable operating characteristics to the main-stream model-based designs. The detailed simulation results are provided in Appendix C.

### 3.3 Sensitivity Analysis of the EI and Cohort Size

The i3+3 design requires the investigators to specify $p_T$ and EI. Our simulation studies have examined the performance of i3+3 with different $p_T$ values. We now assess the performance of the i3+3 design when the length of EI is varied. In addition, we change the cohort size from 3 to different values since the i3+3 design can be applied to any cohort size values despite the name.

*Sensitivity to different EI* Table 3 presents the results of the sensitivity analysis. Specifically, we select a sequence of equal-spaced values of $\varepsilon_1, \varepsilon_2$ so that the resulting symmetric EIs range from $(p_T \pm 0 * p_T)$ to $(p_T \pm 0.2 * p_T)$. For each EI value and each of the 14 scenarios with the same $p_T$, we simulated 1,000 trials with a cohort size of 3 and total sample size of 30. Table 3(a) displays the safety, reliability and percent toxicity (defined as the percentage of patients having DLTs in all of the simulated trials) for different combinations of $p_T$ and EI values. For all three $p_T$ values, safety (the percentage of patients treated at or below the true MTD) and reliability (the probability of identifying the true MTD) tend to increase with the length of EI. The percent toxicity is rarely affected by changing the length of EIs. This is reasonable and a discussion will be provided next.

Sensitivity to different cohort size We fix $p_T = 0.3$ and EI = [0.25, 0.35], and conduct simulations using i3+3 with cohort sizes of 2,3,4,5,6 or a random cohort size within a trial. Since the sample size 4 is not a factor of the total sample size 30, we simulated trials of cohort size 4 with sample size 28 and 32. The results of total sample size 28 are not as good as those with sample size 32. Therefore, we suspect that the larger sample size might have led to better performance. The random value is generated from {2,3,4,5} with equal probability and applied to each step of the decision making within a trial. For each cohort size, 1,000 simulated trials are performed. The results are shown in Table 3(b). The performance of the i3+3 design is robust to different cohort sizes.

## 4 DISCUSSION

The i3+3 design is a simple and intuitive rule-based design demonstrating superior performance than the 3+3 design and similar operating characteristics with the model-based designs. The i3+3 design is versatile since it can accommodate different target $p_T$ values and different cohort sizes. Being a rule-based approach, the i3+3 design can be attractive to clinicians who are used to rule-based designs like 3+3. Other than the slightly different decision tables (Figure A1 in Appendix D), the i3+3, mTPI, and mTPI-2 designs use the same safety rules and same inference to select the MTD. Because the mTPI and mTPI-2 designs have been well established in the literature with desirable performance (4,5,11,12), we decide not to conduct more and larger simulations, such as a simulation study with 1,000 random scenarios (13), to further evaluate the operating characteristics of the i3+3 design. The i3+3 design is expected to perform at a similar level as the mTPI and mTPI-2 designs due to its similar but slightly improved decision table, which puts i3+3 in par with other major model-based designs.

It is interesting that a rule-based design like i3+3 can achieve similar performance as model-based designs. This seems to defy the common statistical belief that model-based inference is superior than rule-based inference. We believe the model-based inference is still superior in general as they offer flexibility and scope for extension. Model-based designs can handle dose-skipping, dose-finding studies with a large number of similar doses, various dose ranges, and more complex patient outcome such as late-onset toxicity. However, in the special case of standard phase I dose-finding trials, due to the ethical constraints to maintain a strict safety control for dosing decisions, model-based inference is heavily restricted by safety rules such as the escalation with over dose control (EWOC, 14) for CRM, the aforementioned Safety rules 1&2 for mTPI, mTPI-2, and BOIN. Bounded by theses practical rules, we believe model-based inference can be approximated by rule-based inference.

In addition, the reason i3+3 performs well is partly due to the use of two values $\frac{x}{n}$ and $\frac{x-1}{n}$ in the decision rules (see Table 1), instead of just one value $\frac{x}{n}$ as seen in the 3+3 design. Let us define the difference $(\frac{x}{n} - \frac{x-1}{n}) = \frac{1}{n}$ the minimum rate difference (MRD) which measures the smallest difference in the observed toxicity rates based on *n* patients. As *n* becomes larger, the MRD becomes smaller, and the variability in the data is smaller. The i3+3 design smartly compares MRD with the length of the EI in order to select between two decisions, de-escalate or stay. It chooses to stay instead of de-escalation when the observed toxicity rate $\frac{x}{n}$ is above the EI and $\frac{x-1}{n}$ is below the EI. This is because the variability of the data is so large that two adjacent data points, $\frac{x}{n}$ and $\frac{x-1}{n}$, reside on either side of the EI, thereby unable to inform whether the true toxicity rate is near or inside the EI. A smart cohort size can be specified based on the comparison between MRD and the EI length. For example, one could choose the cohort size *C* so that the MRD value $\frac{1}{3 \times c}$ with 3 cohorts at a dose is lower than the EI length.

The i3+3 design, like mTPI, mTPI-2, BOIN and CCD, uses data on the current dose to make up-and-down decisions for dose finding. This type of up-and-down decisions generates a Markov chain of dose assignments (Ivanova et al., 2007) which borrows information across doses that have been used for treatment. This is in contrast to the explicit information sharing through a dose-response model as seen in the CRM and BLRM designs. Both philosophies can work well if the goal is to identify the MTD in a classical dose-finding trial. One advantage of model-based approaches is the ability to accommodate prior information such as historical data from previous clinical trials. However, the model-free approach like the i3+3 or model-assisted approaches like mTPI are practically simpler, and offer transparency in the decisions to clinicians.

The i3+3 design only depends on the input of $p_T$ and EI. The target toxicity rate $p_T$ is easily determined as most phase I trials use either 30% or 17%. These two values are results of the popularity of the 3+3 design, which typically targets MTD with toxicity rates in this range. The equivalence interval can be elicited by providing a heuristic boundary around the target $p_T$. Let us consider an example. Suppose $p_T$ is set at 0.3. This means that if 100 patients were treated at the MTD, about 30 patients would experience DLT. Then the clinician can find a number lower than 30, say 25, so that any number below 25 would warrant an escalation as the dose is deemed lower than the MTD. Similarly, a number above 30, say 35, can be elicited so that any number above 35 would warrant a de-escalation as the dose is deemed higher than the MTD. These two numbers give a rate of 0.25 as the lower bound and 0.35 as the higher bound of the EI. That is, this elicitation gives an EI of [0.25, 0.35]. Changing the two numbers results in different EIs. Through the aforementioned sensitivity analysis, we found that the i3+3 design performs well with a wider EI. This is because a wider EI allows a wider range of doses to be considered as the MTD. However, the EI cannot be too wide to become clinically meaningless. For example, an EI of [0.1, 0.6] is not acceptable since it implies any doses with a toxicity rate between 0.1 and 0.6 can be considered as the MTD. We recommend that the EI is in the range of $(p_T \pm \alpha * p_T)$ for $\alpha$ no more than 20% since a >20% deviation from the target value $p_T$ is usually considered excessive in practice.

Users in practice can modify the decision rules and stopping rules based on specific scenario.

The dose assignment decisions of i3+3 must be used with caution when no more than 3 patients are treated at any dose. For example, when 1 out of 1 patient or 2 patients experienced DLT, the i3+3 design recommend to "Stay", which might be considered too risky. However, in practice it is rare to make a dosing decision after 1 or 2 patients are assigned to a dose. Therefore, users can eliminate the decisions corresponding to 1 or 2 patients if such decisions will never be considered.

Alternative stopping rule, such as stop a trial if more than $m$ number of consecutive patients have been assigned to a dose (e.g., $m = 6$ or 9 defined by the users), can be considered. This allows early stopping of a trial, although it may lead to premature stopping due to limited sample size.

Lastly, despite the name i3+3, the proposed design can be used with different cohort size other than 3 as demonstrated in the sensitivity analysis. We name the new design i3+3 to highlight the algorithmic nature of the method just like the 3+3 design.

**Table 1**: The dose-finding algorithm of the i3+3 design. Notation: 1) $d$: the current dose that is being tested in the trial; 2) $n$: the number of patients at the current dose; 3) $x$: the number of patients with DLTs at the current dose. The target toxicity probability is assumed to be $p_T$ and the equivalence interval is denoted as $[p_T - \varepsilon_1, p_T + \varepsilon_2]$. A value is "below" the EI means that the value is smaller than $(p_T - \varepsilon_1)$, the lower bound of the EI. A value is "inside" the EI means that the value is larger than or equal to $(p_T - \varepsilon_1)$ but smaller than or equal to $(p_T + \varepsilon_2)$. A value is "above" the EI mean that the value is larger than $(p_T + \varepsilon_2)$.

| Current dose: $d$,   No. enrolled: $n$,   No. DLTs: $x$ | |
|---|---|
| **Condition** | **Next dose level** |
| $\frac{x}{n}$ below EI | $d+1$ |
| $\frac{x}{n}$ inside EI | $d$ |
| $\frac{x}{n}$ above EI and $\frac{x-1}{n}$ below EI | $d$ |
| $\frac{x}{n}$ above EI and $\frac{x-1}{n}$ inside EI | $d-1$ |
| $\frac{x}{n}$ above EI and $\frac{x-1}{n}$ above EI | $d-1$ |

**Table 2:** Two cases to illustrate the decisions of the i3+3 design in comparison to the mTPI and mTPI-2 designs. In case 1, the target toxicity probability is 0.3, the EI is $[0.25, 0.35]$, and a total of 6 patients are treated at the current dose. In case 2, the target toxicity probability is 0.17, the EI is $[0.12, 0.22]$, and a total of 3 patients are treated at the current dose. Highlighted are where i3+3 is different from mTPI or mTPI-2.

| Current dose: $d$ | | Target probability: $p_T = 0.3$ | |
|---|---|---|---|
| No. enrolled: $n=6$ | | EI: $[p_T-\varepsilon_1, p_T+\varepsilon_2] = [0.25, 0.35]$ | |
| No. DLTs: $x$ | Next dose level | | |
| | i3+3 | mTPI | mTPI-2 |
| 0 | $d+1$ | $d+1$ | $d+1$ |
| 1 | $d+1$ | $d+1$ | $d+1$ |
| 2 | $d$ | $d$ | $d$ |
| 3 | $d-1$ | $d$ | $d-1$ |
| 4 | $d-1$ | $d-1$ | $d-1$ |
| 5 | $d-1$ | $d-1$ | $d-1$ |
| 6 | $d-1$ | $d-1$ | $d-1$ |

Case 1

| Current dose: $d$ | | Target probability: $p_T = 0.17$ | |
|---|---|---|---|
| No. enrolled: $n=3$ | | EI: $[p_T-\varepsilon_1, p_T+\varepsilon_2] = [0.12, 0.22]$ | |
| No. DLTs: $x$ | Next dose level | | |
| | i3+3 | mTPI | mTPI-2 |
| 0 | $d+1$ | $d+1$ | $d+1$ |
| 1 | $d$ | $d$ | $d-1$ |
| 2 | $d-1$ | $d-1$ | $d-1$ |
| 3 | $d-1$ | $d-1$ | $d-1$ |

Case 2

**Table 3**. The operating characteristics of the i3+3 design for varying length of EI (a) and cohort size (b). Safety is defined as the fraction of patients treated at or below the true MTD. Reliability is defined as the fraction of trials selecting the true MTD. %Toxicity is defined as the fraction of patients experiencing in all the simulated trials under each row. The subscripts are standard deviations of the mean values across all the scenarios. Each mean value is averaged across all the simulated trials for a given scenario. In (a), each row corresponds to a different combination of $p_T$ and EI, where EI is denoted by $[p_T-\varepsilon_1, p_T+\varepsilon_2]$. We fix sample size at 30 and cohort size 3, and simulate 1,000 trials for each of the 42 scenarios in Appendix A. Each row corresponds to 14 scenarios with the same $p_T$ value. In (b), we set $p_T = 0.3$ and EI= [0.25, 0.35], using the 14 scenarios in Appendix A. Each row corresponds to a different cohort size. Averaged values and the standard deviations of safety, reliability and percent toxicity are listed. Since the cohort size 4

is not a factor of the sample size 30, trials are simulated with both sample size 28 and sample size 30. 4⁻: Trials simulated with cohort size 4 and sample size 28; 4⁺: Trials simulated with cohort size 4 and sample size 32; **: Random cohort size means randomly selecting the cohort size from the set {2,3,4,5} with replacement and equal probability.

| $p_T$ | EI | Safety | Reliability | %Toxicity |
|---|---|---|---|---|
| 0.1 | 0.1 | $0.411_{0.286}$ | $0.436_{0.232}$ | $0.111_{0.037}$ |
| | [0.095, 0.105] | $0.411_{0.286}$ | $0.439_{0.230}$ | $0.111_{0.037}$ |
| | [0.090, 0.110] | $0.479_{0.321}$ | $0.489_{0.206}$ | $0.111_{0.037}$ |
| | [0.085, 0.115] | $0.470_{0.321}$ | $0.502_{0.229}$ | $0.111_{0.037}$ |
| | [0.080, 0.120] | $0.503_{0.311}$ | $0.549_{0.189}$ | $0.107_{0.037}$ |
| 0.17 | 0.17 | $0.433_{0.305}$ | $0.433_{0.287}$ | $0.164_{0.049}$ |
| | [0.1615, 0.1785] | $0.496_{0.309}$ | $0.428_{0.273}$ | $0.150_{0.046}$ |
| | [0.1530, 0.1870] | $0.537_{0.332}$ | $0.465_{0.251}$ | $0.150_{0.046}$ |
| | [0.1445, 0.1955] | $0.537_{0.332}$ | $0.467_{0.251}$ | $0.150_{0.046}$ |
| | [0.1360, 0.2040] | $0.578_{0.344}$ | $0.565_{0.234}$ | $0.150_{0.046}$ |
| 0.3 | 0.3 | $0.579_{0.308}$ | $0.476_{0.291}$ | $0.246_{0.063}$ |
| | [0.285, 0.315] | $0.613_{0.318}$ | $0.507_{0.272}$ | $0.246_{0.063}$ |
| | [0.270, 0.330] | $0.614_{0.318}$ | $0.548_{0.235}$ | $0.246_{0.062}$ |
| | [0.255, 0.345] | $0.649_{0.334}$ | $0.609_{0.213}$ | $0.245_{0.062}$ |
| | [0.240, 0.360] | $0.731_{0.277}$ | $0.674_{0.199}$ | $0.243_{0.062}$ |

(a)

| Cohort size | Safety | Reliability | % Toxicity |
|---|---|---|---|
| 2 | $0.733_{0.254}$ | $0.631_{0.218}$ | $0.246_{0.062}$ |
| 3 | $0.731_{0.277}$ | $0.657_{0.216}$ | $0.243_{0.062}$ |
| 4⁻ | $0.711_{0.278}$ | $0.638_{0.217}$ | $0.251_{0.060}$ |
| 4⁺ | $0.805_{0.267}$ | $0.572_{0.276}$ | $0.218_{0.071}$ |
| 5 | $0.726_{0.293}$ | $0.633_{0.232}$ | $0.235_{0.072}$ |
| 6 | $0.723_{0.279}$ | $0.640_{0.217}$ | $0.244_{0.062}$ |
| Random** | $0.743_{0.275}$ | $0.630_{0.229}$ | $0.239_{0.064}$ |

(b)

| Number of DLTs \ Number of Patients | 1 | 2 | 3 | 4 | 5 | 6 | 7 | 8 | 9 | 10 | 11 | 12 | 13 | 14 | 15 |
|---|---|---|---|---|---|---|---|---|---|---|---|---|---|---|---|
| 0  | E  | E  | E | E | E | E | E | E | E | E | E | E | E | E | E |
| 1  | S  | S  | S | S | E | E | E | E | E | E | E | E | E | E | E |
| 2  |    | DU | D | D | S | S | S | S | E | E | E | E | E | E | E |
| 3  |    |    | DU| D | D | D | D | D | S | S | S | S | E | E | E |
| 4  |    |    |   | DU| DU| DU| D | D | D | D | D | S | S | S | S |
| 5  |    |    |   |   | DU| DU| DU| DU| DU| D | D | D | D | D | S |
| 6  |    |    |   |   |   | DU| DU| DU| DU| DU| DU| D | D | D | D |
| 7  |    |    |   |   |   |   | DU| DU| DU| DU| DU| DU| DU| D | D |
| 8  |    |    |   |   |   |   |   | DU| DU| DU| DU| DU| DU| DU| DU|
| 9  |    |    |   |   |   |   |   |   | DU| DU| DU| DU| DU| DU| DU|
| 10 |    |    |   |   |   |   |   |   |   | DU| DU| DU| DU| DU| DU|
| 11 |    |    |   |   |   |   |   |   |   |   | DU| DU| DU| DU| DU|
| 12 |    |    |   |   |   |   |   |   |   |   |   | DU| DU| DU| DU|
| 13 |    |    |   |   |   |   |   |   |   |   |   |   | DU| DU| DU|
| 14 |    |    |   |   |   |   |   |   |   |   |   |   |    | DU| DU|
| 15 |    |    |   |   |   |   |   |   |   |   |   |   |    |    | DU|

* **E**: Escalate to the next higher dose; **S**: Stay at the same dose; **D**: De-escalate to the previous lower dose; **DU**: De-escalate to the previous lower dose and the current dose will never be used again in the trial;

**Figure 1:** An example of the decision tables for the i3+3 design. For each "Number of Patients" (column), the decisions "D", "S", "E" are listed for each "Number of DLTs" (row). Letter "DU" stands for de-escalating to the previous lower dose, and the current dose is deemed unacceptable according to Safety Rule 2 so that it will never be used again for the rest of the trial. Here, the target toxicity probability $p_T = 0.3$ and EI $= [0.25, \ 0.35]$.

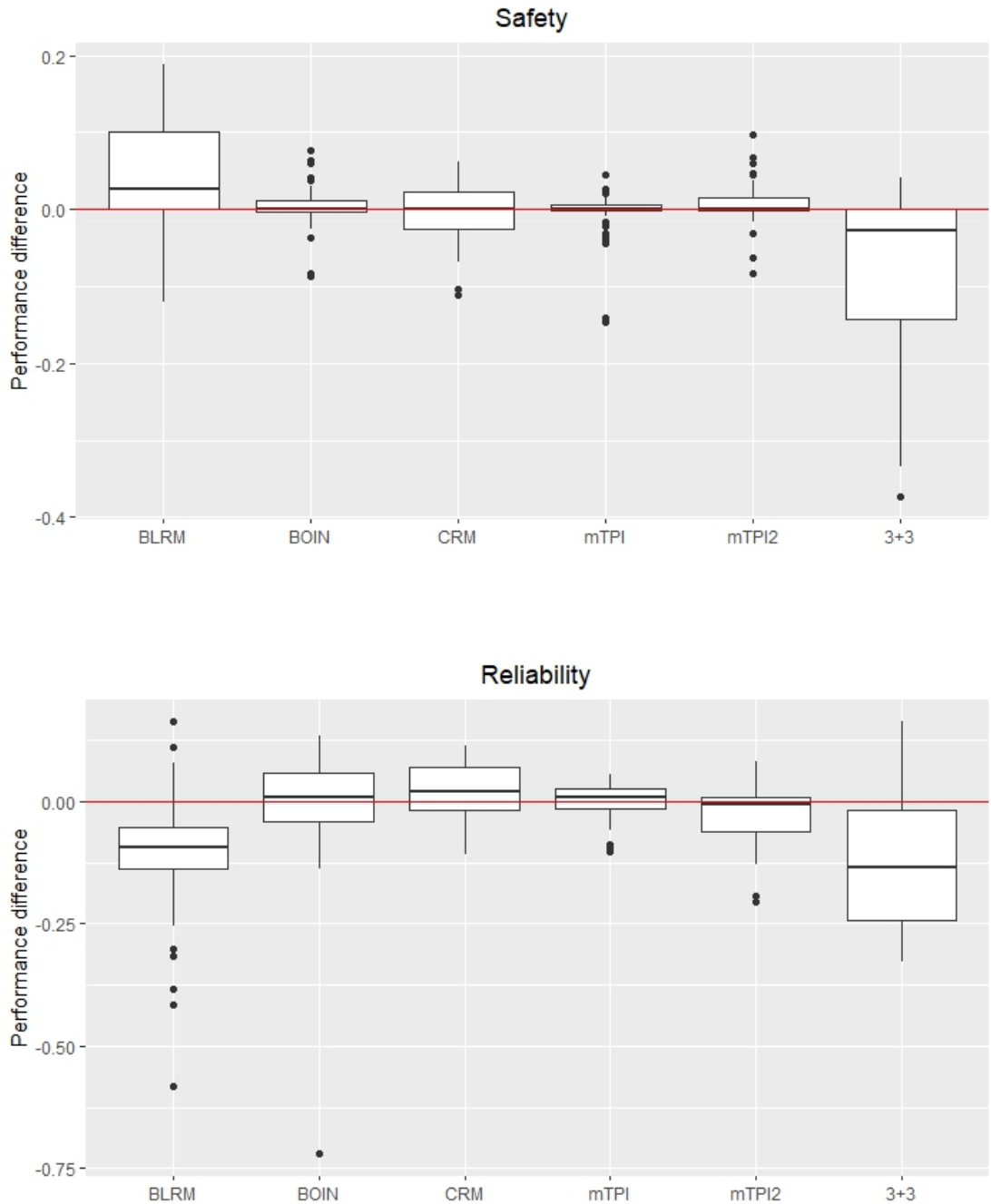

**Figure 2**: Comparison of *safety* and *reliability*. Six designs (mTPI2, mTPI, BOIN, CRM, BLRM and 3+3) are compared with i3+3. Denote any of the six designs as design X. Upper panel [Safety]: each boxplot describes the differences of the safety values between design X and i3+3 averaged across 42 scenarios. A value greater than zero means design X puts more percentages of patients at the MTD or doses below the MTD than i3+3. Lower panel [Reliability]: each boxplot describes the differences of the reliability values between design X and i3+3 averaged across all 42 scenarios. A value greater than zero means design X is more likely to identify the true MTD than i3+3.

# Appendix A
# Simulation scenarios

We adopted the 42 scenarios in Ji and Wang (2013) for $p_T = 0.1, 0.17,$ and $0.3$, where scenarios for $p_T = 0.17$ were generated by subtracting 0.03 from the original scenarios for $p_T = 0.2$ in Ji and Wang (2013). Each scenario contains six true toxicity probabilities for all the six doses in the simulated clinical trials. They are listed as follows.

$p_T = 0.1$

| Scenarios# | Dose 1 | Dose 2 | Dose 3 | Dose 4 | Dose 5 | Dose 6 |
|---|---|---|---|---|---|---|
| 1 | 0.04 | 0.05 | 0.06 | 0.07 | 0.08 | 0.09 |
| 2 | 0.15 | 0.2 | 0.25 | 0.3 | 0.35 | 0.4 |
| 3 | 0.01 | 0.1 | 0.2 | 0.25 | 0.3 | 0.35 |
| 4 | 0.01 | 0.02 | 0.03 | 0.04 | 0.1 | 0.25 |
| 5 | 0.05 | 0.4 | 0.5 | 0.6 | 0.65 | 0.7 |
| 6 | 0.01 | 0.03 | 0.05 | 0.4 | 0.5 | 0.6 |
| 7 | 0.01 | 0.02 | 0.03 | 0.04 | 0.05 | 0.4 |
| 8 | 0.09 | 0.11 | 0.13 | 0.15 | 0.17 | 0.19 |
| 9 | 0.05 | 0.07 | 0.09 | 0.11 | 0.13 | 0.15 |
| 10 | 0.01 | 0.03 | 0.05 | 0.07 | 0.09 | 0.11 |
| 11 | 0.02 | 0.04 | 0.08 | 0.12 | 0.17 | 0.25 |
| 12 | 0.02 | 0.04 | 0.07 | 0.1 | 0.15 | 0.2 |
| 13 | 0.1 | 0.15 | 0.2 | 0.25 | 0.3 | 0.35 |
| 14 | 0.01 | 0.03 | 0.05 | 0.06 | 0.08 | 0.1 |

$p_T = 0.17$

| Scenarios# | Dose 1 | Dose 2 | Dose 3 | Dose 4 | Dose 5 | Dose 6 |
|---|---|---|---|---|---|---|
| 15 | 0 | 0.02 | 0.05 | 0.08 | 0.11 | 0.14 |
| 16 | 0.22 | 0.32 | 0.37 | 0.47 | 0.57 | 0.67 |
| 17 | 0 | 0.17 | 0.37 | 0.57 | 0.77 | 0.92 |
| 18 | 0.01 | 0.03 | 0.05 | 0.07 | 0.17 | 0.47 |
| 19 | 0.02 | 0.47 | 0.77 | 0.87 | 0.92 | 0.96 |
| 20 | 0 | 0.02 | 0.07 | 0.47 | 0.67 | 0.87 |
| 21 | 0 | 0 | 0.04 | 0.07 | 0.12 | 0.67 |
| 22 | 0.16 | 0.18 | 0.2 | 0.22 | 0.24 | 0.26 |
| 23 | 0.12 | 0.14 | 0.16 | 0.18 | 0.2 | 0.22 |
| 24 | 0.08 | 0.1 | 0.12 | 0.14 | 0.16 | 0.18 |
| 25 | 0.02 | 0.08 | 0.14 | 0.2 | 0.26 | 0.32 |
| 26 | 0.02 | 0.07 | 0.12 | 0.17 | 0.27 | 0.34 |
| 27 | 0.17 | 0.22 | 0.27 | 0.32 | 0.37 | 0.42 |
| 28 | 0.02 | 0.05 | 0.08 | 0.11 | 0.14 | 0.17 |

$p_T = 0.3$

| Scenarios# | Dose 1 | Dose 2 | Dose 3 | Dose 4 | Dose 5 | Dose 6 |
|---|---|---|---|---|---|---|
| 29 | 0.02 | 0.05 | 0.1 | 0.15 | 0.2 | 0.25 |
| 30 | 0.35 | 0.45 | 0.5 | 0.6 | 0.7 | 0.8 |
| 31 | 0.01 | 0.3 | 0.55 | 0.65 | 0.8 | 0.95 |
| 32 | 0.04 | 0.06 | 0.08 | 0.1 | 0.3 | 0.6 |
| 33 | 0.05 | 0.6 | 0.8 | 0.9 | 0.95 | 0.99 |
| 34 | 0.01 | 0.05 | 0.1 | 0.6 | 0.7 | 0.9 |

| 35 | 0.01 | 0.03 | 0.07 | 0.1  | 0.15 | 0.75 |
| 36 | 0.29 | 0.31 | 0.33 | 0.35 | 0.37 | 0.39 |
| 37 | 0.25 | 0.27 | 0.29 | 0.31 | 0.33 | 0.35 |
| 38 | 0.21 | 0.23 | 0.25 | 0.27 | 0.29 | 0.31 |
| 39 | 0.05 | 0.2  | 0.27 | 0.33 | 0.39 | 0.45 |
| 40 | 0.05 | 0.1  | 0.2  | 0.3  | 0.4  | 0.4  |
| 41 | 0.3  | 0.35 | 0.4  | 0.45 | 0.5  | 0.55 |
| 42 | 0.15 | 0.18 | 0.21 | 0.24 | 0.27 | 0.3  |

## Appendix B

### Details of the CRM and BLRM design implemented in simulation study.

*The Continuous Reassessment Method (CRM)*

In the CRM design, we used the one-parameter power model

$$p_i = c_i^\theta \text{ for } \theta > 0, i = 1, \ldots I,$$

where $\theta$ is the parameter of the model, $p_i$ is the toxicity rate at dose $i$, and $c_i$'s are pre-specified prior toxicity probability ('skeleton') values, monotonically increasing with $i$. The skeleton $c_i$'s reflect the initial guesses of toxicity probability at each dose. We adopted the method proposed in Lee and Cheung (2011) for selecting the skeleton based on indifference intervals. We used half-width $\lambda = 0.05$ and we used log-normal prior $\pi(\theta) \sim lognormal(0, 1.34)$.

The posterior distribution of θ is given by

$$p(\theta|n, y) \propto \prod_{i=1}^{I} (c_i^\theta)^{y_i} (1 - c_i^\theta)^{n_i - y_i} \pi(\theta),$$

where $n = \{n_1, \ldots, n_I\}$ are the number patients, and $y = \{y_1, \ldots, y_I\}$ are the number of patients experienced DLTs.

The CRM design starts the trial by treating the first cohort at the lowest dose $d_1$. The posterior distribution of π(θ) is updated based on the accumulating information across all dose levels, and the posterior mean of π(θ) is used as the point estimate of $\theta$ to guide dose assignment. In particular, the next cohort will be assigned to the dose with $\hat{p}_i$ closest to $p_T$. In our simulation, an additional no-skipping escalation rule is added in order to ensure the safety of the trial conduct. In addition, the Safety rules 1 and 2 as stated in Section 2.1 are also applied in CRM design. Lastly, no escalation is permitted if the empirical DLT rate of the most recent cohort is higher than $p_T$.

*The Bayesian Logistic Regression Method (BLRM)*

BLRM assumes a two-parameter logistic model

$$logit(p_i) = log(\alpha) + \beta \log\left(\frac{d_i}{d_{ref}}\right)$$

where $\alpha, \beta$ are unknown parameters, $d_i$ is the raw dosage at dose level $i$, and $d_{ref}$ is the reference dose level. Model parameters α and β follow a multivariate log-normal prior,

$$\binom{\alpha}{\beta} \sim lognormal\left(\binom{\mu_1}{\mu_2}, \Sigma\right), \text{ where } \Sigma = \begin{pmatrix} \sigma_1^2 & \rho\sigma_1\sigma_2 \\ \rho\sigma_1\sigma_2 & \sigma_2^2 \end{pmatrix}$$

Let $\theta = \{\mu 1, \mu 2, \sigma 1, \sigma 2, \rho\}$ be the hyperparameter set of the model. In the simulation, we used a default set of doses, $d_i = 5 \times i$, and the default reference dose level is the ceiling of $\frac{(I+1)}{2}$, where $I$ is the number of doses. In addition, we used the quantile-based non-informative prior calculator proposed by Neuenschwander et al. (2008) to obtain the values of $\theta$, and use the same prior guess for the lowest and highest doses as described in Appendix A.1 in Neuenschwander et al. (2008). The posterior distribution of $(\alpha, \beta)$ is given

$$p(\theta|n, y) \propto \prod_{i=1}^{I} (c_i^\theta)^{y_i} (1 - c_i^\theta)^{n_i - y_i} \pi(\theta)$$

where $n = \{n_1, \dots, n_I\}$ are the number patients, and $y = \{y_1, \dots, y_I\}$ are the number of patients experienced DLTs.

The BLRM starts a trial by treating the first cohort at the lowest patient. After each cohort is treated, the standard dose recommendation of BLRM relies on maximizing the probability of targeted toxicity interval $(p_T + \varepsilon_1, p_T + \varepsilon_2)$. In our simulation, we used $\varepsilon_1 = \varepsilon_2 = 0.05$. In particular, the next cohort will be assigned to the dose whose posterior probability in the targeted toxicity interval is the largest, i.e. $\underset{i=1,\ldots,I}{argmax} \Pr(p_i \in (p_T + \varepsilon_1, p_T + \varepsilon_2)|y,n)$. The BLRM imposes an overdose control rule (Escalation with Overdose Control, EWOC) such that the probability of excessive toxicity of the recommended dose should be less than a given threshold $p_{EWOC}$, i.e. $\Pr(p_i > p_T + \varepsilon_2|y,n) < p_{EWOC}$. In our simulation, we used $p_{EWOC} = 0.5$. In addition, the Safety rules 1 and 2 as stated in Section 2.1 are also applied in BLRM design. Besides safety rule I, if all doses violate the EWOC rule, the trial will be terminated before the maximum the sample size is reached.

# Appendix C
## Operating characteristics comparing i3+3 with other phase I designs

Six designs (mTPI2, mTPI, BOIN, CRM, BLRM and 3+3) are compared with i3+3. A sample size of 30 patients is used for each trial with a cohort size of 3. For the mTPI and mTPI-2 designs, the EI is $[p_T - 0.05, p_T + 0.05]$. For the BOIN design, we set $(p_T - \lambda_e) = 0.05$ and $(\lambda_d - p_T) = 0.05$ so that the actual intervals for decision making between BOIN and mTPI or mTPI-2 are identical. For the BLRM design, the EWOC parameter is set at 0.5. Under each simulation scenario, we generate 1,000 simulated trials.

## Scenario 1

| Dose Level | Target Toxicity Prob. = 0.1 True Tox Prob. | Selection Prob. | | | | | | | # of Patients Treated | | | | | | | # of Toxicities | | | | | | |
|---|---|---|---|---|---|---|---|---|---|---|---|---|---|---|---|---|---|---|---|---|---|---|
| | | i3+3 | mTPI | mTPI-2 | 3+3 | BOIN | BLRM | CRM | i3+3 | mTPI | mTPI-2 | 3+3 | BOIN | BLRM | CRM | i3+3 | mTPI | mTPI-2 | 3+3 | BOIN | BLRM | CRM |
| 1 | 0.04 | 0.092 | 0.084 | 0.107 | 0.019 | 0.109 | 0.077 | 0.076 | 6.261 | 6.261 | 7.05 | 3.372 | 6.888 | 5.766 | 6.102 | 0.256 | 0.249 | 0.277 | 0.133 | 0.275 | 0.224 | 0.259 |
| 2 | 0.05 | 0.123 | 0.119 | 0.168 | 0.035 | 0.135 | 0.235 | 0.168 | 5.916 | 5.943 | 6.081 | 3.432 | 5.64 | 9.123 | 6.474 | 0.337 | 0.32 | 0.334 | 0.168 | 0.274 | 0.469 | 0.356 |
| 3 | 0.06 | 0.123 | 0.204 | 0.157 | 0.051 | 0.129 | 0.362 | 0.171 | 5.022 | 5.4 | 5.055 | 3.537 | 4.791 | 10.011 | 5.403 | 0.288 | 0.343 | 0.322 | 0.223 | 0.274 | 0.637 | 0.33 |
| 4 | 0.07 | 0.218 | 0.145 | 0.162 | 0.059 | 0.174 | 0.217 | 0.172 | 4.134 | 3.969 | 3.831 | 3.444 | 4.065 | 2.694 | 4.248 | 0.311 | 0.289 | 0.276 | 0.256 | 0.259 | 0.197 | 0.305 |
| 5 | 0.08 | 0.135 | 0.132 | 0.125 | 0.065 | 0.141 | 0.023 | 0.122 | 2.787 | 2.856 | 2.979 | 3.324 | 3.378 | 0.18 | 3.042 | 0.196 | 0.225 | 0.23 | 0.271 | 0.286 | 0.017 | 0.243 |
| 6 | 0.09 | 0.286 | 0.291 | 0.256 | 0.75 | 0.296 | 0 | 0.266 | 5.376 | 5.01 | 4.443 | 3.003 | 4.878 | 0.003 | 4.14 | 0.465 | 0.458 | 0.409 | 0.279 | 0.411 | 0 | 0.371 |
| | | i3+3 | mTPI | mTPI-2 | 3+3 | BOIN | BLRM | CRM | | | | | | | | | | | | | | |
| Prob. of Select MTD | | 0.885 | 0.891 | 0.868 | 0.96 | 0.875 | 0.837 | 0.899 | | | | | | | | | | | | | | |
| Prob. of Toxicity | | 0.063 | 0.064 | 0.063 | 0.066 | 0.06 | 0.056 | 0.063 | | | | | | | | | | | | | | |
| Prob. of Select Dose-over-MTD | | 0 | 0 | 0 | 0 | 0 | 0 | 0 | | | | | | | | | | | | | | |
| Prob. of No Selection | | 0.023 | 0.025 | 0.025 | 0.021 | 0.016 | 0.086 | 0.025 | | | | | | | | | | | | | | |

## Scenario 2

| Dose Level | Target Toxicity Prob. = 0.1 True Tox Prob. | Selection Prob. | | | | | | | # of Patients Treated | | | | | | | # of Toxicities | | | | | | |
|---|---|---|---|---|---|---|---|---|---|---|---|---|---|---|---|---|---|---|---|---|---|---|
| | | i3+3 | mTPI | mTPI-2 | 3+3 | BOIN | BLRM | CRM | i3+3 | mTPI | mTPI-2 | 3+3 | BOIN | BLRM | CRM | i3+3 | mTPI | mTPI-2 | 3+3 | BOIN | BLRM | CRM |
| | | | | | | | | | 29.439 | | | | 29.64 | | 29.409 | | | | | | | |
| 1 | 0.15 | 0.356 | 0.345 | 0.371 | 0.265 | 0.437 | 0.276 | 0.386 | 12.417 | 12.939 | 14.649 | 4.611 | 14.403 | 10.017 | 13.743 | 1.912 | 1.969 | 2.219 | 0.684 | 2.089 | 1.554 | 2.082 |
| 2 | 0.2 | 0.138 | 0.145 | 0.119 | 0.233 | 0.144 | 0.104 | 0.162 | 6.339 | 6.144 | 5.058 | 3.921 | 5.526 | 5.739 | 5.793 | 1.203 | 1.221 | 1.01 | 0.812 | 1.08 | 1.216 | 1.148 |
| 3 | 0.25 | 0.039 | 0.074 | 0.037 | 0.153 | 0.041 | 0.021 | 0.054 | 2.784 | 2.652 | 2.088 | 2.73 | 2.22 | 1.326 | 2.403 | 0.732 | 0.637 | 0.507 | 0.698 | 0.534 | 0.322 | 0.584 |
| 4 | 0.3 | 0.014 | 0.009 | 0.007 | 0.103 | 0.012 | 0.001 | 0.01 | 0.963 | 0.963 | 0.807 | 1.662 | 0.741 | 0.084 | 0.756 | 0.277 | 0.289 | 0.246 | 0.495 | 0.227 | 0.027 | 0.219 |
| 5 | 0.35 | 0.003 | 0.003 | 0.001 | 0.034 | 0.005 | 0 | 0 | 0.267 | 0.231 | 0.21 | 0.816 | 0.243 | 0 | 0.171 | 0.102 | 0.087 | 0.081 | 0.307 | 0.071 | 0 | 0.069 |
| 6 | 0.4 | 0 | 0 | 0 | 0.017 | 0 | 0 | 0 | 0.069 | 0.03 | 0.039 | 0.24 | 0.057 | 0 | 0.027 | 0.025 | 0.016 | 0.018 | 0.099 | 0.025 | 0 | 0.014 |
| | | i3+3 | mTPI | mTPI-2 | 3+3 | BOIN | BLRM | CRM | | | | | | | | | | | | | | |
| Prob. of Select MTD | | 0.356 | 0.345 | 0.371 | 0.265 | 0.437 | 0.276 | 0.386 | | | | | | | | | | | | | | |
| Prob. of Toxicity | | 0.186 | 0.184 | 0.179 | 0.221 | 0.174 | 0.182 | 0.18 | | | | | | | | | | | | | | |
| Prob. of Select Dose-over-MTD | | 0.194 | 0.231 | 0.164 | 0.54 | 0.202 | 0.126 | 0.226 | | | | | | | | | | | | | | |
| Prob. of No Selection | | 0.45 | 0.424 | 0.465 | 0.195 | 0.361 | 0.598 | 0.388 | | | | | | | | | | | | | | |

## Scenario 3

| Dose Level | Target Toxicity Prob. = 0.1 True Tox Prob. | Selection Prob. | | | | | | | # of Patients Treated | | | | | | | # of Toxicities | | | | | | |
|---|---|---|---|---|---|---|---|---|---|---|---|---|---|---|---|---|---|---|---|---|---|---|
| | | i3+3 | mTPI | mTPI-2 | 3+3 | BOIN | BLRM | CRM | i3+3 | mTPI | mTPI-2 | 3+3 | BOIN | BLRM | CRM | i3+3 | mTPI | mTPI-2 | 3+3 | BOIN | BLRM | CRM |
| 1 | 0.01 | 0.265 | 0.195 | 0.273 | 0.093 | 0.2 | 0.194 | 0.166 | 7.38 | 6.246 | 8.295 | 3.336 | 8.247 | 8.085 | 7.047 | 0.087 | 0.066 | 0.085 | 0.024 | 0.089 | 0.084 | 0.075 |
| 2 | 0.1 | 0.446 | 0.497 | 0.526 | 0.284 | 0.523 | 0.608 | 0.538 | 11.109 | 12.348 | 12.186 | 4.524 | 12.177 | 14.781 | 12.618 | 1.134 | 1.225 | 1.225 | 0.453 | 1.157 | 1.438 | 1.266 |
| 3 | 0.2 | 0.184 | 0.213 | 0.137 | 0.242 | 0.196 | 0.157 | 0.224 | 6.894 | 7.152 | 5.904 | 4.365 | 6.027 | 5.874 | 6.678 | 1.33 | 1.447 | 1.197 | 0.89 | 1.173 | 1.173 | 1.353 |
| 4 | 0.25 | 0.076 | 0.07 | 0.044 | 0.197 | 0.061 | 0.01 | 0.056 | 3.054 | 2.745 | 2.301 | 3.069 | 2.394 | 0.429 | 2.514 | 0.727 | 0.694 | 0.6 | 0.741 | 0.59 | 0.101 | 0.636 |
| 5 | 0.3 | 0.021 | 0.02 | 0.014 | 0.109 | 0.019 | 0 | 0.009 | 1.11 | 1.107 | 0.942 | 2.01 | 0.852 | 0.012 | 0.804 | 0.322 | 0.314 | 0.255 | 0.615 | 0.241 | 0.002 | 0.226 |
| 6 | 0.35 | 0.002 | 0.002 | 0.002 | 0.074 | 0 | 0 | 0.004 | 0.396 | 0.327 | 0.297 | 0.813 | 0.279 | 0 | 0.264 | 0.13 | 0.124 | 0.112 | 0.294 | 0.094 | 0 | 0.098 |
| | | i3+3 | mTPI | mTPI-2 | 3+3 | BOIN | BLRM | CRM | | | | | | | | | | | | | | |
| Prob. of Select MTD | | 0.446 | 0.497 | 0.526 | 0.284 | 0.523 | 0.608 | 0.538 | | | | | | | | | | | | | | |
| Prob. of Toxicity | | 0.125 | 0.129 | 0.116 | 0.167 | 0.112 | 0.096 | 0.122 | | | | | | | | | | | | | | |
| Prob. of Select Dose-over-MTD | | 0.283 | 0.305 | 0.197 | 0.622 | 0.276 | 0.167 | 0.293 | | | | | | | | | | | | | | |
| Prob. of No Selection | | 0.006 | 0.003 | 0.004 | 0.001 | 0.001 | 0.031 | 0.003 | | | | | | | | | | | | | | |

## Scenario 4

| Dose Level | Target Toxicity Prob. = 0.1 True Tox Prob. | Selection Prob. | | | | | | | # of Patients Treated | | | | | | | # of Toxicities | | | | | | |
|---|---|---|---|---|---|---|---|---|---|---|---|---|---|---|---|---|---|---|---|---|---|---|
| | | i3+3 | mTPI | mTPI-2 | 3+3 | BOIN | BLRM | CRM | i3+3 | mTPI | mTPI-2 | 3+3 | BOIN | BLRM | CRM | i3+3 | mTPI | mTPI-2 | 3+3 | BOIN | BLRM | CRM |
| 1 | 0.01 | 0.006 | 0.015 | 0.016 | 0.004 | 0.023 | 0.004 | 0.009 | 3.648 | 3.858 | 4.089 | 3.099 | 4.272 | 3.786 | 3.714 | 0.036 | 0.042 | 0.045 | 0.034 | 0.052 | 0.034 | 0.039 |
| 2 | 0.02 | 0.036 | 0.022 | 0.076 | 0.011 | 0.066 | 0.101 | 0.044 | 4.335 | 4.404 | 4.647 | 3.207 | 4.581 | 7.719 | 4.425 | 0.089 | 0.091 | 0.093 | 0.067 | 0.102 | 0.143 | 0.095 |
| 3 | 0.03 | 0.068 | 0.144 | 0.115 | 0.014 | 0.093 | 0.346 | 0.109 | 4.554 | 4.764 | 4.785 | 3.282 | 4.581 | 12.066 | 5.007 | 0.134 | 0.125 | 0.13 | 0.1 | 0.123 | 0.359 | 0.135 |
| 4 | 0.04 | 0.343 | 0.245 | 0.299 | 0.091 | 0.243 | 0.472 | 0.309 | 5.307 | 5.34 | 5.811 | 3.528 | 5.646 | 5.316 | 6.732 | 0.217 | 0.218 | 0.229 | 0.142 | 0.223 | 0.214 | 0.271 |
| 5 | 0.1 | 0.466 | 0.48 | 0.422 | 0.367 | 0.457 | 0.049 | 0.423 | 7.269 | 6.966 | 6.858 | 4.518 | 6.948 | 0.42 | 6.729 | 0.7 | 0.717 | 0.701 | 0.439 | 0.642 | 0.046 | 0.674 |
| 6 | 0.25 | 0.079 | 0.091 | 0.069 | 0.51 | 0.117 | 0.003 | 0.103 | 4.848 | 4.593 | 3.735 | 3.804 | 3.948 | 0.024 | 3.318 | 1.22 | 1.167 | 0.977 | 0.975 | 1.009 | 0.002 | 0.882 |
| | | i3+3 | mTPI | mTPI-2 | 3+3 | BOIN | BLRM | CRM | | | | | | | | | | | | | | |
| Prob. of Select MTD | | 0.466 | 0.48 | 0.422 | 0.367 | 0.457 | 0.049 | 0.423 | | | | | | | | | | | | | | |
| Prob. of Toxicity | | 0.08 | 0.079 | 0.073 | 0.082 | 0.0718 | 0.027 | 0.07 | | | | | | | | | | | | | | |
| Prob. of Select Dose-over-MTD | | 0.079 | 0.091 | 0.069 | 0.51 | 0.117 | 0.003 | 0.103 | | | | | | | | | | | | | | |
| Prob. of No Selection | | 0.002 | 0.003 | 0.003 | 0.003 | 0.001 | 0.025 | 0.003 | | | | | | | | | | | | | | |

## Scenario 5

| Dose Level | Target Toxicity Prob. = 0.1 True Tox Prob. | Selection Prob. | | | | | | | # of Patients Treated | | | | | | | # of Toxicities | | | | | | |
|---|---|---|---|---|---|---|---|---|---|---|---|---|---|---|---|---|---|---|---|---|---|---|
| | | i3+3 | mTPI | mTPI-2 | 3+3 | BOIN | BLRM | CRM | i3+3 | mTPI | mTPI-2 | 3+3 | BOIN | BLRM | CRM | i3+3 | mTPI | mTPI-2 | 3+3 | BOIN | BLRM | CRM |
| 1 | 0.05 | 0.937 | 0.939 | 0.937 | 0.742 | 0.956 | 0.681 | 0.918 | 22.308 | 22.38 | 22.926 | 5.394 | 23.004 | 17.952 | 23.037 | 1.11 | 1.14 | 1.152 | 0.277 | 1.13 | 0.931 | 1.172 |
| 2 | 0.4 | 0.01 | 0.007 | 0.009 | 0.197 | 0.015 | 0.025 | 0.039 | 5.724 | 6.033 | 5.439 | 4.737 | 5.67 | 4.986 | 5.46 | 2.29 | 2.408 | 2.194 | 1.924 | 2.229 | 2.035 | 2.188 |
| 3 | 0.5 | 0.001 | 0.001 | 0.001 | 0.028 | 0.003 | 0 | 0 | 0.975 | 0.648 | 0.612 | 1.314 | 0.738 | 0.411 | 0.567 | 0.504 | 0.352 | 0.332 | 0.654 | 0.357 | 0.185 | 0.31 |
| 4 | 0.6 | 0 | 0 | 0 | 0.002 | 0 | 0 | 0 | 0.099 | 0.057 | 0.051 | 0.225 | 0.084 | 0 | 0.054 | 0.056 | 0.038 | 0.036 | 0.139 | 0.057 | 0 | 0.037 |
| 5 | 0.65 | 0 | 0 | 0 | 0.001 | 0 | 0 | 0 | 0.006 | 0 | 0 | 0.021 | 0.003 | 0 | 0 | 0.003 | 0 | 0 | 0.011 | 0.002 | 0 | 0 |
| 6 | 0.7 | 0 | 0 | 0 | 0 | 0 | 0 | 0 | 0 | 0 | 0 | 0.003 | 0 | 0 | 0 | 0 | 0 | 0 | 0.002 | 0 | 0 | 0 |
| | | i3+3 | mTPI | mTPI-2 | 3+3 | BOIN | BLRM | CRM | | | | | | | | | | | | | | |
| Prob. of Select MTD | | 0.937 | 0.939 | 0.937 | 0.742 | 0.956 | 0.681 | 0.918 | | | | | | | | | | | | | | |
| Prob. of Toxicity | | 0.136 | 0.135 | 0.128 | 0.257 | 0.128 | 0.135 | 0.127 | | | | | | | | | | | | | | |
| Prob. of Select Dose-over-MTD | | 0.011 | 0.008 | 0.01 | 0.228 | 0.018 | 0.025 | 0.039 | | | | | | | | | | | | | | |
| Prob. of No Selection | | 0.052 | 0.053 | 0.053 | 0.03 | 0.026 | 0.294 | 0.043 | | | | | | | | | | | | | | |

### Scenario 6

Target Toxicity Prob. = 0.1

| Dose Level | True Tox Prob. | Selection Prob. | | | | | | | # of Patients Treated | | | | | | | # of Toxicities | | | | | | |
|---|---|---|---|---|---|---|---|---|---|---|---|---|---|---|---|---|---|---|---|---|---|---|
| | | i3+3 | mTPI | mTPI-2 | 3+3 | BOIN | BLRM | CRM | i3+3 | mTPI | mTPI-2 | 3+3 | BOIN | BLRM | CRM | i3+3 | mTPI | mTPI-2 | 3+3 | BOIN | BLRM | CRM |
| 1 | 0.01 | 0.014 | 0.017 | 0.031 | 0.013 | 0.036 | 0.013 | 0.015 | 3.762 | 3.984 | 4.452 | 3.132 | 4.647 | 4.11 | 3.912 | 0.032 | 0.044 | 0.05 | 0.033 | 0.055 | 0.038 | 0.044 |
| 2 | 0.03 | 0.122 | 0.089 | 0.177 | 0.024 | 0.132 | 0.186 | 0.148 | 5.472 | 5.457 | 6.279 | 3.342 | 6.093 | 9.135 | 6.915 | 0.164 | 0.161 | 0.19 | 0.11 | 0.19 | 0.251 | 0.209 |
| 3 | 0.05 | 0.828 | 0.868 | 0.768 | 0.739 | 0.792 | 0.718 | 0.809 | 14.832 | 14.943 | 14.076 | 5.31 | 14.103 | 14.103 | 14.076 | 0.759 | 0.741 | 0.695 | 0.257 | 0.696 | 0.702 | 0.708 |
| 4 | 0.4 | 0.036 | 0.023 | 0.021 | 0.194 | 0.038 | 0.058 | 0.025 | 4.998 | 4.791 | 4.416 | 4.725 | 4.464 | 1.956 | 4.407 | 1.964 | 1.919 | 1.758 | 1.884 | 1.758 | 0.78 | 1.77 |
| 5 | 0.5 | 0 | 0 | 0 | 0.023 | 0.001 | 0 | 0 | 0.855 | 0.669 | 0.621 | 1.29 | 0.594 | 0.027 | 0.537 | 0.431 | 0.332 | 0.31 | 0.673 | 0.279 | 0.011 | 0.261 |
| 6 | 0.6 | 0 | 0 | 0 | 0.005 | 0 | 0 | 0 | 0.081 | 0.081 | 0.081 | 0.183 | 0.075 | 0 | 0.078 | 0.051 | 0.05 | 0.05 | 0.111 | 0.039 | 0 | 0.048 |

| | i3+3 | mTPI | mTPI-2 | 3+3 | BOIN | BLRM | CRM |
|---|---|---|---|---|---|---|---|
| Prob. of Select MTD | 0.828 | 0.868 | 0.768 | 0.739 | 0.792 | 0.718 | 0.809 |
| Prob. of Toxicity | 0.113 | 0.109 | 0.102 | 0.171 | 0.1 | 0.061 | 0.102 |
| Prob. of Select Dose-over-MTD | 0.036 | 0.023 | 0.021 | 0.222 | 0.039 | 0.058 | 0.025 |
| Prob. of No Selection | 0 | 0.003 | 0.003 | 0.002 | 0.001 | 0.025 | 0.003 |

### Scenario 7

Target Toxicity Prob. = 0.1

| Dose Level | True Tox Prob. | Selection Prob. | | | | | | | # of Patients Treated | | | | | | | # of Toxicities | | | | | | |
|---|---|---|---|---|---|---|---|---|---|---|---|---|---|---|---|---|---|---|---|---|---|---|
| | | i3+3 | mTPI | mTPI-2 | 3+3 | BOIN | BLRM | CRM | i3+3 | mTPI | mTPI-2 | 3+3 | BOIN | BLRM | CRM | i3+3 | mTPI | mTPI-2 | 3+3 | BOIN | BLRM | CRM |
| 1 | 0.01 | 0.011 | 0.015 | 0.016 | 0.004 | 0.023 | 0.004 | 0.009 | 3.603 | 3.858 | 4.086 | 3.102 | 4.272 | 3.786 | 3.714 | 0.026 | 0.042 | 0.045 | 0.033 | 0.052 | 0.034 | 0.039 |
| 2 | 0.02 | 0.025 | 0.022 | 0.075 | 0.008 | 0.066 | 0.101 | 0.044 | 4.434 | 4.404 | 4.581 | 3.225 | 4.581 | 7.719 | 4.425 | 0.104 | 0.091 | 0.092 | 0.075 | 0.102 | 0.143 | 0.095 |
| 3 | 0.03 | 0.063 | 0.142 | 0.108 | 0.025 | 0.091 | 0.346 | 0.108 | 4.692 | 4.764 | 4.731 | 3.315 | 4.512 | 12.054 | 4.935 | 0.14 | 0.125 | 0.128 | 0.096 | 0.123 | 0.359 | 0.13 |
| 4 | 0.04 | 0.252 | 0.144 | 0.177 | 0.029 | 0.145 | 0.464 | 0.26 | 4.68 | 4.434 | 4.716 | 3.435 | 4.473 | 5.304 | 6.477 | 0.183 | 0.184 | 0.2 | 0.166 | 0.167 | 0.213 | 0.276 |
| 5 | 0.05 | 0.639 | 0.663 | 0.604 | 0.642 | 0.646 | 0.057 | 0.53 | 8.805 | 8.787 | 8.418 | 4.971 | 8.55 | 0.441 | 7.368 | 0.439 | 0.464 | 0.429 | 0.243 | 0.42 | 0.025 | 0.364 |
| 6 | 0.4 | 0.01 | 0.011 | 0.017 | 0.29 | 0.028 | 0.003 | 0.046 | 3.786 | 3.678 | 3.336 | 4.038 | 3.588 | 0.027 | 3.006 | 1.525 | 1.488 | 1.369 | 1.615 | 1.405 | 0.006 | 1.253 |

| | i3+3 | mTPI | mTPI-2 | 3+3 | BOIN | BLRM | CRM |
|---|---|---|---|---|---|---|---|
| Prob. of Select MTD | 0.639 | 0.663 | 0.604 | 0.642 | 0.646 | 0.057 | 0.53 |
| Prob. of Toxicity | 0.081 | 0.08 | 0.076 | 0.101 | 0.076 | 0.027 | 0.072 |
| Prob. of Select Dose-over-MTD | 0.01 | 0.011 | 0.017 | 0.29 | 0.028 | 0.003 | 0.046 |
| Prob. of No Selection | 0 | 0.003 | 0.003 | 0.002 | 0.001 | 0.025 | 0.003 |

### Scenario 8

Target Toxicity Prob. = 0.1

| Dose Level | True Tox Prob. | Selection Prob. | | | | | | | # of Patients Treated | | | | | | | # of Toxicities | | | | | | |
|---|---|---|---|---|---|---|---|---|---|---|---|---|---|---|---|---|---|---|---|---|---|---|
| | | i3+3 | mTPI | mTPI-2 | 3+3 | BOIN | BLRM | CRM | i3+3 | mTPI | mTPI-2 | 3+3 | BOIN | BLRM | CRM | i3+3 | mTPI | mTPI-2 | 3+3 | BOIN | BLRM | CRM |
| 1 | 0.09 | 0.277 | 0.273 | 0.323 | 0.1 | 0.303 | 0.182 | 0.26 | 10.461 | 10.29 | 11.736 | 3.981 | 11.331 | 7.842 | 10.344 | 0.948 | 0.95 | 1.075 | 0.39 | 1.04 | 0.704 | 0.957 |
| 2 | 0.11 | 0.188 | 0.188 | 0.195 | 0.13 | 0.198 | 0.317 | 0.247 | 6.435 | 7.056 | 6.468 | 3.774 | 6.405 | 9.36 | 7.335 | 0.713 | 0.801 | 0.747 | 0.406 | 0.668 | 1.002 | 0.839 |
| 3 | 0.13 | 0.149 | 0.179 | 0.144 | 0.121 | 0.149 | 0.181 | 0.163 | 4.659 | 4.656 | 4.317 | 3.447 | 4.293 | 5.202 | 4.668 | 0.594 | 0.596 | 0.543 | 0.474 | 0.526 | 0.673 | 0.589 |
| 4 | 0.15 | 0.143 | 0.103 | 0.096 | 0.137 | 0.106 | 0.033 | 0.107 | 2.994 | 2.76 | 2.496 | 3 | 2.652 | 0.765 | 2.805 | 0.436 | 0.41 | 0.372 | 0.442 | 0.395 | 0.12 | 0.411 |
| 5 | 0.17 | 0.069 | 0.074 | 0.057 | 0.111 | 0.061 | 0.004 | 0.067 | 1.542 | 1.533 | 1.47 | 2.523 | 1.554 | 0.042 | 1.485 | 0.275 | 0.256 | 0.237 | 0.446 | 0.237 | 0.003 | 0.24 |
| 6 | 0.19 | 0.029 | 0.046 | 0.031 | 0.306 | 0.05 | 0 | 0.041 | 1.338 | 1.329 | 1.071 | 1.752 | 1.095 | 0 | 0.981 | 0.251 | 0.244 | 0.202 | 0.338 | 0.184 | 0 | 0.188 |

| | i3+3 | mTPI | mTPI-2 | 3+3 | BOIN | BLRM | CRM |
|---|---|---|---|---|---|---|---|
| Prob. of Select MTD | 0.757 | 0.743 | 0.758 | 0.488 | 0.756 | 0.713 | 0.777 |
| Prob. of Toxicity | 0.117 | 0.118 | 0.115 | 0.135 | 0.112 | 0.108 | 0.117 |
| Prob. of Select Dose-over-MTD | 0.098 | 0.12 | 0.088 | 0.417 | 0.366 | 0.004 | 0.108 |
| Prob. of No Selection | 0.145 | 0.137 | 0.154 | 0.095 | 0.133 | 0.283 | 0.115 |

### Scenario 9

Target Toxicity Prob. = 0.1

| Dose Level | True Tox Prob. | Selection Prob. | | | | | | | # of Patients Treated | | | | | | | # of Toxicities | | | | | | |
|---|---|---|---|---|---|---|---|---|---|---|---|---|---|---|---|---|---|---|---|---|---|---|
| | | i3+3 | mTPI | mTPI-2 | 3+3 | BOIN | BLRM | CRM | i3+3 | mTPI | mTPI-2 | 3+3 | BOIN | BLRM | CRM | i3+3 | mTPI | mTPI-2 | 3+3 | BOIN | BLRM | CRM |
| 1 | 0.05 | 0.16 | 0.149 | 0.186 | 0.047 | 0.174 | 0.097 | 0.128 | 7.446 | 7.584 | 8.721 | 3.537 | 8.61 | 6.285 | 7.326 | 0.364 | 0.379 | 0.432 | 0.18 | 0.415 | 0.328 | 0.366 |
| 2 | 0.07 | 0.171 | 0.173 | 0.229 | 0.071 | 0.221 | 0.317 | 0.229 | 6.516 | 6.945 | 6.969 | 3.63 | 6.942 | 10.161 | 7.5 | 0.495 | 0.519 | 0.513 | 0.244 | 0.489 | 0.678 | 0.551 |
| 3 | 0.09 | 0.164 | 0.26 | 0.202 | 0.11 | 0.177 | 0.329 | 0.239 | 5.511 | 5.69 | 5.208 | 3.75 | 5.34 | 8.346 | 5.757 | 0.478 | 0.483 | 0.468 | 0.346 | 0.467 | 0.769 | 0.497 |
| 4 | 0.11 | 0.225 | 0.146 | 0.158 | 0.111 | 0.181 | 0.11 | 0.149 | 4.116 | 4.059 | 3.789 | 3.543 | 3.834 | 1.647 | 4.215 | 0.432 | 0.493 | 0.448 | 0.42 | 0.441 | 0.167 | 0.513 |
| 5 | 0.13 | 0.149 | 0.119 | 0.096 | 0.122 | 0.104 | 0.01 | 0.114 | 2.943 | 2.418 | 2.391 | 3.18 | 2.427 | 0.111 | 2.427 | 0.379 | 0.328 | 0.329 | 0.409 | 0.315 | 0.023 | 0.323 |
| 6 | 0.15 | 0.094 | 0.12 | 0.094 | 0.513 | 0.119 | 0 | 0.109 | 2.76 | 2.634 | 2.181 | 2.514 | 2.346 | 0 | 2.034 | 0.437 | 0.381 | 0.313 | 0.382 | 0.334 | 0 | 0.295 |

| | i3+3 | mTPI | mTPI-2 | 3+3 | BOIN | BLRM | CRM |
|---|---|---|---|---|---|---|---|
| Prob. of Select MTD | 0.963 | 0.967 | 0.965 | 0.974 | 0.976 | 0.863 | 0.968 |
| Prob. of Toxicity | 0.088 | 0.088 | 0.086 | 0.098 | 0.083 | 0.074 | 0.087 |
| Prob. of Select Dose-over-MTD | 0 | 0 | 0 | 0 | 0 | 0 | 0 |
| Prob. of No Selection | 0.037 | 0.033 | 0.035 | 0.026 | 0.024 | 0.137 | 0.032 |

### Scenario 10

Target Toxicity Prob. = 0.1

| Dose Level | True Tox Prob. | Selection Prob. | | | | | | | # of Patients Treated | | | | | | | # of Toxicities | | | | | | |
|---|---|---|---|---|---|---|---|---|---|---|---|---|---|---|---|---|---|---|---|---|---|---|
| | | i3+3 | mTPI | mTPI-2 | 3+3 | BOIN | BLRM | CRM | i3+3 | mTPI | mTPI-2 | 3+3 | BOIN | BLRM | CRM | i3+3 | mTPI | mTPI-2 | 3+3 | BOIN | BLRM | CRM |
| 1 | 0.01 | 0.021 | 0.017 | 0.025 | 0.014 | 0.035 | 0.013 | 0.011 | 4.041 | 3.984 | 4.404 | 3.129 | 4.629 | 4.11 | 3.879 | 0.049 | 0.044 | 0.05 | 0.031 | 0.055 | 0.038 | 0.044 |
| 2 | 0.03 | 0.082 | 0.069 | 0.145 | 0.018 | 0.112 | 0.179 | 0.086 | 5.169 | 5.394 | 5.781 | 3.306 | 5.643 | 9 | 5.421 | 0.164 | 0.158 | 0.168 | 0.105 | 0.18 | 0.247 | 0.158 |
| 3 | 0.05 | 0.143 | 0.223 | 0.182 | 0.049 | 0.162 | 0.421 | 0.193 | 5.724 | 5.853 | 5.772 | 3.489 | 5.643 | 11.862 | 6.072 | 0.283 | 0.288 | 0.281 | 0.168 | 0.272 | 0.595 | 0.301 |
| 4 | 0.07 | 0.292 | 0.176 | 0.197 | 0.076 | 0.178 | 0.326 | 0.221 | 5.301 | 5.01 | 4.902 | 3.609 | 4.695 | 4.092 | 5.592 | 0.397 | 0.361 | 0.354 | 0.251 | 0.321 | 0.259 | 0.396 |
| 5 | 0.09 | 0.212 | 0.219 | 0.215 | 0.094 | 0.203 | 0.035 | 0.224 | 4.134 | 4.035 | 4.224 | 3.615 | 4.194 | 0.252 | 4.392 | 0.385 | 0.351 | 0.372 | 0.327 | 0.338 | 0.021 | 0.386 |
| 6 | 0.11 | 0.248 | 0.293 | 0.233 | 0.748 | 0.309 | 0.001 | 0.262 | 5.58 | 5.649 | 4.842 | 3.207 | 5.172 | 0.015 | 4.569 | 0.669 | 0.672 | 0.587 | 0.36 | 0.54 | 0.001 | 0.554 |

| | i3+3 | mTPI | mTPI-2 | 3+3 | BOIN | BLRM | CRM |
|---|---|---|---|---|---|---|---|
| Prob. of Select MTD | 0.895 | 0.911 | 0.827 | 0.967 | 0.852 | 0.783 | 0.9 |
| Prob. of Toxicity | 0.065 | 0.063 | 0.061 | 0.061 | 0.057 | 0.04 | 0.061 |
| Prob. of Select Dose-over-MTD | 0 | 0 | 0 | 0 | 0 | 0 | 0 |
| Prob. of No Selection | 0.002 | 0.003 | 0.003 | 0.001 | 0.001 | 0.025 | 0.003 |

## Scenario 11

| Dose Level | True Tox Prob. | Selection Prob. | | | | | | | # of Patients Treated | | | | | | | # of Toxicities | | | | | | |
|---|---|---|---|---|---|---|---|---|---|---|---|---|---|---|---|---|---|---|---|---|---|---|
| Target Toxicity Prob. = 0.1 | | i3+3 | mTPI | mTPI-2 | 3+3 | BOIN | BLRM | CRM | i3+3 | mTPI | mTPI-2 | 3+3 | BOIN | BLRM | CRM | i3+3 | mTPI | mTPI-2 | 3+3 | BOIN | BLRM | CRM |
| 1 | 0.02 | 0.052 | 0.03 | 0.043 | 0.012 | 0.063 | 0.041 | 0.023 | 4.728 | 4.566 | 5.178 | 3.183 | 5.613 | 4.854 | 4.431 | 0.101 | 0.087 | 0.099 | 0.058 | 0.119 | 0.103 | 0.086 |
| 2 | 0.04 | 0.169 | 0.146 | 0.238 | 0.071 | 0.193 | 0.282 | 0.179 | 6.357 | 6.573 | 7.158 | 3.525 | 6.912 | 10.56 | 6.882 | 0.286 | 0.273 | 0.275 | 0.15 | 0.272 | 0.417 | 0.284 |
| 3 | 0.08 | 0.265 | 0.374 | 0.334 | 0.134 | 0.259 | 0.457 | 0.331 | 7.152 | 7.578 | 7.422 | 3.915 | 6.822 | 10.908 | 7.932 | 0.556 | 0.579 | 0.568 | 0.327 | 0.543 | 0.864 | 0.629 |
| 4 | 0.12 | 0.312 | 0.223 | 0.218 | 0.193 | 0.27 | 0.154 | 0.272 | 5.817 | 5.547 | 5.214 | 3.975 | 5.304 | 2.061 | 6.051 | 0.722 | 0.706 | 0.674 | 0.502 | 0.581 | 0.258 | 0.764 |
| 5 | 0.17 | 0.151 | 0.174 | 0.126 | 0.232 | 0.162 | 0.01 | 0.146 | 3.429 | 3.498 | 3.156 | 3.669 | 3.462 | 0.12 | 3.024 | 0.595 | 0.593 | 0.529 | 0.639 | 0.584 | 0.019 | 0.509 |
| 6 | 0.25 | 0.043 | 0.045 | 0.033 | 0.351 | 0.048 | 0 | 0.041 | 2.313 | 2.052 | 1.686 | 2.472 | 1.773 | 0 | 1.494 | 0.544 | 0.481 | 0.404 | 0.623 | 0.429 | 0 | 0.36 |
| | | i3+3 | mTPI | mTPI-2 | 3+3 | BOIN | BLRM | CRM | | | | | | | | | | | | | | |
| Prob. of Select MTD | | 0.577 | 0.597 | 0.552 | 0.327 | 0.529 | 0.611 | 0.603 | | | | | | | | | | | | | | |
| Prob. of Toxicity | | 0.094 | 0.091 | 0.085 | 0.111 | 0.085 | 0.058 | 0.088 | | | | | | | | | | | | | | |
| Prob. of Select Dose-over-MTD | | 0.194 | 0.219 | 0.159 | 0.583 | 0.21 | 0.01 | 0.187 | | | | | | | | | | | | | | |
| Prob. of No Selection | | 0.008 | 0.008 | 0.008 | 0.007 | 0.005 | 0.056 | 0.008 | | | | | | | | | | | | | | |

## Scenario 12

| Dose Level | True Tox Prob. | Selection Prob. | | | | | | | # of Patients Treated | | | | | | | # of Toxicities | | | | | | |
|---|---|---|---|---|---|---|---|---|---|---|---|---|---|---|---|---|---|---|---|---|---|---|
| Target Toxicity Prob. = 0.1 | | i3+3 | mTPI | mTPI-2 | 3+3 | BOIN | BLRM | CRM | i3+3 | mTPI | mTPI-2 | 3+3 | BOIN | BLRM | CRM | i3+3 | mTPI | mTPI-2 | 3+3 | BOIN | BLRM | CRM |
| 1 | 0.02 | 0.042 | 0.029 | 0.041 | 0.017 | 0.063 | 0.04 | 0.021 | 4.632 | 4.554 | 5.154 | 3.216 | 5.607 | 4.827 | 4.554 | 0.085 | 0.087 | 0.099 | 0.063 | 0.119 | 0.1 | 0.085 |
| 2 | 0.04 | 0.136 | 0.12 | 0.206 | 0.05 | 0.164 | 0.242 | 0.155 | 5.91 | 6.234 | 6.696 | 3.456 | 6.393 | 10.137 | 6.474 | 0.264 | 0.257 | 0.261 | 0.146 | 0.258 | 0.404 | 0.268 |
| 3 | 0.07 | 0.234 | 0.332 | 0.295 | 0.086 | 0.214 | 0.443 | 0.287 | 6.582 | 7.071 | 6.978 | 3.726 | 6.288 | 10.914 | 7.425 | 0.435 | 0.46 | 0.454 | 0.268 | 0.426 | 0.738 | 0.5 |
| 4 | 0.1 | 0.341 | 0.234 | 0.238 | 0.158 | 0.282 | 0.2 | 0.288 | 5.958 | 5.532 | 5.334 | 3.864 | 5.367 | 2.451 | 6.135 | 0.635 | 0.603 | 0.593 | 0.397 | 0.477 | 0.236 | 0.664 |
| 5 | 0.15 | 0.165 | 0.186 | 0.144 | 0.206 | 0.174 | 0.019 | 0.156 | 3.909 | 3.735 | 3.408 | 3.81 | 3.783 | 0.171 | 3.333 | 0.605 | 0.573 | 0.52 | 0.555 | 0.565 | 0.023 | 0.518 |
| 6 | 0.2 | 0.076 | 0.091 | 0.068 | 0.476 | 0.098 | 0 | 0.085 | 2.865 | 2.688 | 2.244 | 2.859 | 2.448 | 0.003 | 2.04 | 0.567 | 0.527 | 0.446 | 0.601 | 0.459 | 0 | 0.399 |
| | | i3+3 | mTPI | mTPI-2 | 3+3 | BOIN | BLRM | CRM | | | | | | | | | | | | | | |
| Prob. of Select MTD | | 0.74 | 0.752 | 0.677 | 0.45 | 0.67 | 0.662 | 0.731 | | | | | | | | | | | | | | |
| Prob. of Toxicity | | 0.087 | 0.084 | 0.08 | 0.097 | 0.077 | 0.053 | 0.082 | | | | | | | | | | | | | | |
| Prob. of Select Dose-over-MTD | | 0.076 | 0.091 | 0.068 | 0.476 | 0.098 | 0 | 0.085 | | | | | | | | | | | | | | |
| Prob. of No Selection | | 0.006 | 0.008 | 0.008 | 0.007 | 0.005 | 0.056 | 0.008 | | | | | | | | | | | | | | |

## Scenario 13

| Dose Level | True Tox Prob. | Selection Prob. | | | | | | | # of Patients Treated | | | | | | | # of Toxicities | | | | | | |
|---|---|---|---|---|---|---|---|---|---|---|---|---|---|---|---|---|---|---|---|---|---|---|
| Target Toxicity Prob. = 0.1 | | i3+3 | mTPI | mTPI-2 | 3+3 | BOIN | BLRM | CRM | i3+3 | mTPI | mTPI-2 | 3+3 | BOIN | BLRM | CRM | i3+3 | mTPI | mTPI-2 | 3+3 | BOIN | BLRM | CRM |
| 1 | 0.1 | 0.385 | 0.378 | 0.442 | 0.176 | 0.42 | 0.297 | 0.376 | 11.769 | 11.838 | 14.001 | 4.173 | 13.614 | 9.834 | 12.312 | 1.21 | 1.215 | 1.409 | 0.422 | 1.4 | 0.979 | 1.269 |
| 2 | 0.15 | 0.262 | 0.247 | 0.213 | 0.227 | 0.241 | 0.282 | 0.3 | 8.172 | 8.361 | 7.26 | 4.152 | 7.326 | 8.856 | 8.424 | 1.224 | 1.233 | 1.109 | 0.63 | 1.042 | 1.335 | 1.234 |
| 3 | 0.2 | 0.092 | 0.141 | 0.093 | 0.187 | 0.102 | 0.074 | 0.125 | 4.254 | 4.095 | 3.438 | 3.522 | 3.483 | 3.243 | 3.912 | 0.885 | 0.812 | 0.665 | 0.736 | 0.673 | 0.606 | 0.81 |
| 4 | 0.25 | 0.032 | 0.038 | 0.03 | 0.179 | 0.049 | 0.01 | 0.025 | 1.617 | 1.839 | 1.476 | 2.55 | 1.494 | 0.291 | 1.566 | 0.434 | 0.458 | 0.359 | 0.594 | 0.36 | 0.065 | 0.389 |
| 5 | 0.3 | 0.013 | 0.012 | 0.01 | 0.073 | 0.008 | 0 | 0.009 | 0.684 | 0.591 | 0.519 | 1.563 | 0.537 | 0.015 | 0.462 | 0.18 | 0.172 | 0.154 | 0.528 | 0.154 | 0.005 | 0.134 |
| 6 | 0.35 | 0.003 | 0.001 | 0.001 | 0.055 | 0 | 0 | 0.003 | 0.231 | 0.183 | 0.153 | 0.579 | 0.159 | 0 | 0.141 | 0.08 | 0.061 | 0.046 | 0.199 | 0.052 | 0 | 0.042 |
| | | i3+3 | mTPI | mTPI-2 | 3+3 | BOIN | BLRM | CRM | | | | | | | | | | | | | | |
| Prob. of Select MTD | | 0.647 | 0.625 | 0.655 | 0.403 | 0.661 | 0.579 | 0.676 | | | | | | | | | | | | | | |
| Prob. of Toxicity | | 0.15 | 0.147 | 0.139 | 0.188 | 0.138 | 0.134 | 0.145 | | | | | | | | | | | | | | |
| Prob. of Select Dose-over-MTD | | 0.14 | 0.192 | 0.134 | 0.494 | 0.159 | 0.084 | 0.162 | | | | | | | | | | | | | | |
| Prob. of No Selection | | 0.213 | 0.183 | 0.211 | 0.103 | 0.18 | 0.337 | 0.162 | | | | | | | | | | | | | | |

## Scenario 14

| Dose Level | True Tox Prob. | Selection Prob. | | | | | | | # of Patients Treated | | | | | | | # of Toxicities | | | | | | |
|---|---|---|---|---|---|---|---|---|---|---|---|---|---|---|---|---|---|---|---|---|---|---|
| Target Toxicity Prob. = 0.1 | | i3+3 | mTPI | mTPI-2 | 3+3 | BOIN | BLRM | CRM | i3+3 | mTPI | mTPI-2 | 3+3 | BOIN | BLRM | CRM | i3+3 | mTPI | mTPI-2 | 3+3 | BOIN | BLRM | CRM |
| 1 | 0.01 | 0.014 | 0.017 | 0.025 | 0.014 | 0.035 | 0.013 | 0.011 | 3.789 | 3.984 | 4.404 | 3.132 | 4.629 | 4.11 | 3.879 | 0.033 | 0.044 | 0.05 | 0.033 | 0.055 | 0.038 | 0.044 |
| 2 | 0.03 | 0.071 | 0.068 | 0.143 | 0.023 | 0.112 | 0.179 | 0.085 | 5.136 | 5.379 | 5.742 | 3.312 | 5.634 | 8.997 | 5.394 | 0.151 | 0.157 | 0.167 | 0.101 | 0.18 | 0.247 | 0.157 |
| 3 | 0.05 | 0.101 | 0.202 | 0.154 | 0.033 | 0.15 | 0.403 | 0.167 | 5.4 | 5.616 | 5.472 | 3.447 | 5.391 | 11.769 | 5.793 | 0.257 | 0.273 | 0.266 | 0.175 | 0.261 | 0.588 | 0.284 |
| 4 | 0.06 | 0.299 | 0.172 | 0.19 | 0.058 | 0.161 | 0.341 | 0.216 | 5.226 | 4.818 | 4.782 | 3.486 | 4.47 | 4.161 | 5.412 | 0.297 | 0.293 | 0.285 | 0.204 | 0.266 | 0.221 | 0.324 |
| 5 | 0.08 | 0.22 | 0.201 | 0.208 | 0.078 | 0.19 | 0.037 | 0.212 | 4.26 | 4.05 | 4.257 | 3.567 | 4.137 | 0.276 | 4.467 | 0.327 | 0.328 | 0.34 | 0.288 | 0.301 | 0.023 | 0.355 |
| 6 | 0.1 | 0.295 | 0.337 | 0.277 | 0.792 | 0.351 | 0.002 | 0.306 | 6.189 | 6.078 | 5.268 | 3.252 | 5.715 | 0.018 | 4.98 | 0.661 | 0.638 | 0.566 | 0.327 | 0.54 | 0.001 | 0.534 |
| | | i3+3 | mTPI | mTPI-2 | 3+3 | BOIN | BLRM | CRM | | | | | | | | | | | | | | |
| Prob. of Select MTD | | 0.915 | 0.912 | 0.829 | 0.961 | 0.852 | 0.783 | 0.901 | | | | | | | | | | | | | | |
| Prob. of Toxicity | | 0.058 | 0.058 | 0.056 | 0.056 | 0.053 | 0.038 | 0.057 | | | | | | | | | | | | | | |
| Prob. of Select Dose-over-MTD | | 0 | 0 | 0 | 0 | 0 | 0 | 0 | | | | | | | | | | | | | | |
| Prob. of No Selection | | 0 | 0.003 | 0.003 | 0.002 | 0.001 | 0.025 | 0.003 | | | | | | | | | | | | | | |

## Scenario 1

| Target Toxicity Prob. = 0.17 | | Selection Prob. | | | | | | # of Patients Treated | | | | | | | # of Toxicities | | | | | | |
|---|---|---|---|---|---|---|---|---|---|---|---|---|---|---|---|---|---|---|---|---|---|
| Dose Level | True Tox Prob. | i3+3 | mTPI | mTPI-2 | 3+3 | BOIN | BLRM | CRM | i3+3 | mTPI | mTPI-2 | 3+3 | BOIN | BLRM | CRM | i3+3 | mTPI | mTPI-2 | 3+3 | BOIN | BLRM | CRM |
| 1 | 0.01 | 0.003 | 0.001 | 0.001 | 0.012 | 0.002 | 0.001 | 0.001 | 3.081 | 3.33 | 3.507 | 3.105 | 3.267 | 3.3 | 3.249 | 0 | 0.034 | 0.035 | 0.027 | 0 | 0.036 | 0.034 |
| 2 | 0.02 | 0.015 | 0.015 | 0.028 | 0.015 | 0.015 | 0.017 | 0.015 | 3.606 | 3.759 | 4.428 | 3.216 | 4.149 | 4.656 | 3.606 | 0.064 | 0.069 | 0.086 | 0.075 | 0.08 | 0.096 | 0.068 |
| 3 | 0.05 | 0.066 | 0.058 | 0.121 | 0.068 | 0.064 | 0.17 | 0.059 | 4.638 | 5.085 | 5.829 | 3.492 | 5.337 | 9.126 | 4.701 | 0.227 | 0.279 | 0.299 | 0.147 | 0.261 | 0.45 | 0.254 |
| 4 | 0.08 | 0.16 | 0.212 | 0.239 | 0.109 | 0.184 | 0.361 | 0.181 | 5.643 | 5.604 | 5.736 | 3.822 | 5.913 | 7.776 | 6.195 | 0.47 | 0.418 | 0.441 | 0.327 | 0.469 | 0.636 | 0.463 |
| 5 | 0.11 | 0.262 | 0.194 | 0.237 | 0.136 | 0.251 | 0.25 | 0.268 | 5.145 | 4.713 | 4.881 | 3.732 | 4.887 | 3.276 | 5.373 | 0.584 | 0.491 | 0.527 | 0.414 | 0.533 | 0.351 | 0.563 |
| 6 | 0.14 | 0.494 | 0.52 | 0.374 | 0.658 | 0.484 | 0.192 | 0.476 | 7.887 | 7.509 | 5.619 | 3.168 | 6.447 | 1.623 | 6.876 | 1.098 | 1.105 | 0.837 | 0.464 | 0.822 | 0.209 | 1.006 |
| | | i3+3 | mTPI | mTPI-2 | 3+3 | BOIN | BLRM | CRM | | | | | | | | | | | | | | |
| Prob. of Select MTD | | 0.494 | 0.52 | 0.374 | 0.658 | 0.484 | 0.192 | 0.476 | | | | | | | | | | | | | | |
| Prob. of Toxicity | | 0.081 | 0.08 | 0.074 | 0.071 | 0.072 | 0.06 | 0.08 | | | | | | | | | | | | | | |
| Prob. of Select Dose-over-MTD | | 0 | 0 | 0 | 0 | 0 | 0 | 0 | | | | | | | | | | | | | | |
| Prob. of No Selection | | 0 | 0 | 0 | 0.002 | 0 | 0.009 | 0 | | | | | | | | | | | | | | |

## Scenario 2

| Target Toxicity Prob. = 0.17 | | Selection Prob. | | | | | | | # of Patients Treated | | | | | | | # of Toxicities | | | | | | |
|---|---|---|---|---|---|---|---|---|---|---|---|---|---|---|---|---|---|---|---|---|---|---|
| Dose Level | True Tox Prob. | i3+3 | mTPI | mTPI-2 | 3+3 | BOIN | BLRM | CRM | i3+3 | mTPI | mTPI-2 | 3+3 | BOIN | BLRM | CRM | i3+3 | mTPI | mTPI-2 | 3+3 | BOIN | BLRM | CRM |
| 1 | 0.22 | 0.386 | 0.44 | 0.393 | 0.37 | 0.506 | 0.305 | 0.5 | 16.035 | 15.411 | 17.79 | 5.061 | 17.469 | 11.472 | 16.134 | 3.614 | 3.404 | 3.944 | 1.125 | 3.799 | 2.535 | 3.568 |
| 2 | 0.32 | 0.076 | 0.074 | 0.056 | 0.149 | 0.114 | 0.081 | 0.098 | 5.037 | 5.919 | 4.017 | 3.237 | 4.545 | 4.971 | 5.004 | 1.635 | 1.894 | 1.268 | 1.081 | 1.441 | 1.598 | 1.617 |
| 3 | 0.37 | 0.018 | 0.019 | 0.012 | 0.078 | 0.011 | 0.009 | 0.021 | 1.389 | 1.524 | 0.951 | 1.461 | 1.182 | 1.131 | 1.551 | 0.528 | 0.55 | 0.336 | 0.529 | 0.427 | 0.397 | 0.546 |
| 4 | 0.47 | 0.001 | 0 | 0 | 0.011 | 0.001 | 0.001 | 0 | 0.288 | 0.264 | 0.171 | 0.51 | 0.231 | 0.129 | 0.234 | 0.125 | 0.14 | 0.097 | 0.258 | 0.109 | 0.043 | 0.128 |
| 5 | 0.57 | 0 | 0 | 0 | 0.001 | 0 | 0 | 0 | 0.033 | 0.036 | 0.027 | 0.06 | 0.042 | 0.003 | 0.033 | 0.021 | 0.018 | 0.013 | 0.035 | 0.019 | 0.002 | 0.016 |
| 6 | 0.67 | 0 | 0 | 0 | 0.001 | 0 | 0 | 0 | 0 | 0.003 | 0.003 | 0.009 | 0 | 0 | 0.003 | 0 | 0.002 | 0.002 | 0.004 | 0 | 0 | 0.002 |
| | | i3+3 | mTPI | mTPI-2 | 3+3 | BOIN | BLRM | CRM | | | | | | | | | | | | | | |
| Prob. of Select MTD | | 0.386 | 0.44 | 0.393 | 0.37 | 0.506 | 0.305 | 0.5 | | | | | | | | | | | | | | |
| Prob. of Toxicity | | 0.26 | 0.259 | 0.247 | 0.293 | 0.247 | 0.258 | 0.256 | | | | | | | | | | | | | | |
| Prob. of Select Dose-over-MTD | | 0.095 | 0.093 | 0.068 | 0.24 | 0.126 | 0.091 | 0.119 | | | | | | | | | | | | | | |
| Prob. of No Selection | | 0.519 | 0.467 | 0.539 | 0.39 | 0.368 | 0.604 | 0.381 | | | | | | | | | | | | | | |

## Scenario 3

| Target Toxicity Prob. = 0.17 | | Selection Prob. | | | | | | | # of Patients Treated | | | | | | | # of Toxicities | | | | | | |
|---|---|---|---|---|---|---|---|---|---|---|---|---|---|---|---|---|---|---|---|---|---|---|
| Dose Level | True Tox Prob. | i3+3 | mTPI | mTPI-2 | 3+3 | BOIN | BLRM | CRM | i3+3 | mTPI | mTPI-2 | 3+3 | BOIN | BLRM | CRM | i3+3 | mTPI | mTPI-2 | 3+3 | BOIN | BLRM | CRM |
| 1 | 0.01 | 0.356 | 0.293 | 0.362 | 0.255 | 0.166 | 0.245 | 0.213 | 7.65 | 6.363 | 9.372 | 3.828 | 9.249 | 8.1 | 7.848 | 0 | 0.065 | 0.101 | 0.039 | 0 | 0.083 | 0.079 |
| 2 | 0.17 | 0.592 | 0.648 | 0.608 | 0.545 | 0.725 | 0.669 | 0.692 | 15.402 | 16.503 | 15.768 | 5.334 | 15.138 | 16.107 | 15.372 | 2.621 | 2.763 | 2.639 | 0.88 | 2.526 | 2.709 | 2.595 |
| 3 | 0.37 | 0.052 | 0.059 | 0.03 | 0.186 | 0.109 | 0.076 | 0.095 | 5.979 | 6.414 | 4.341 | 3.9 | 4.974 | 5.148 | 6.036 | 2.233 | 2.398 | 1.631 | 1.507 | 1.799 | 1.871 | 2.257 |
| 4 | 0.57 | 0 | 0 | 0 | 0.013 | 0 | 0 | 0 | 0.9 | 0.669 | 0.477 | 1.11 | 0.615 | 0.387 | 0.705 | 0.489 | 0.385 | 0.277 | 0.653 | 0.376 | 0.233 | 0.404 |
| 5 | 0.77 | 0 | 0 | 0 | 0 | 0 | 0 | 0 | 0.069 | 0.051 | 0.042 | 0.111 | 0.024 | 0.009 | 0.039 | 0.052 | 0.043 | 0.036 | 0.086 | 0.02 | 0.009 | 0.033 |
| 6 | 0.92 | 0 | 0 | 0 | 0 | 0 | 0 | 0 | 0 | 0 | 0 | 0.003 | 0 | 0 | 0 | 0 | 0 | 0 | 0.003 | 0 | 0 | 0 |
| | | i3+3 | mTPI | mTPI-2 | 3+3 | BOIN | BLRM | CRM | | | | | | | | | | | | | | |
| Prob. of Select MTD | | 0.592 | 0.648 | 0.608 | 0.545 | 0.725 | 0.669 | 0.692 | | | | | | | | | | | | | | |
| Prob. of Toxicity | | 0.18 | 0.188 | 0.156 | 0.222 | 0.157 | 0.165 | 0.179 | | | | | | | | | | | | | | |
| Prob. of Select Dose-over-MTD | | 0.052 | 0.059 | 0.03 | 0.199 | 0.109 | 0.076 | 0.095 | | | | | | | | | | | | | | |
| Prob. of No Selection | | 0 | 0 | 0 | 0.001 | 0 | 0.01 | 0 | | | | | | | | | | | | | | |

## Scenario 4

| Target Toxicity Prob. = 0.17 | | Selection Prob. | | | | | | | # of Patients Treated | | | | | | | # of Toxicities | | | | | | |
|---|---|---|---|---|---|---|---|---|---|---|---|---|---|---|---|---|---|---|---|---|---|---|
| Dose Level | True Tox Prob. | i3+3 | mTPI | mTPI-2 | 3+3 | BOIN | BLRM | CRM | i3+3 | mTPI | mTPI-2 | 3+3 | BOIN | BLRM | CRM | i3+3 | mTPI | mTPI-2 | 3+3 | BOIN | BLRM | CRM |
| 1 | 0.01 | 0.001 | 0.001 | 0.001 | 0.008 | 0.002 | 0.003 | 0.001 | 3.246 | 3.33 | 3.594 | 3.102 | 3.552 | 3.363 | 3.273 | 0.03 | 0.034 | 0.036 | 0.029 | 0.025 | 0.036 | 0.034 |
| 2 | 0.03 | 0.018 | 0.02 | 0.035 | 0.031 | 0.022 | 0.024 | 0.016 | 4.026 | 4.05 | 4.686 | 3.387 | 4.428 | 5.016 | 3.75 | 0.128 | 0.103 | 0.122 | 0.117 | 0.13 | 0.152 | 0.098 |
| 3 | 0.05 | 0.049 | 0.054 | 0.112 | 0.054 | 0.069 | 0.161 | 0.064 | 4.278 | 4.977 | 5.502 | 3.519 | 5.1 | 8.835 | 4.716 | 0.193 | 0.269 | 0.279 | 0.188 | 0.25 | 0.429 | 0.25 |
| 4 | 0.07 | 0.364 | 0.363 | 0.395 | 0.246 | 0.307 | 0.46 | 0.32 | 6.177 | 5.751 | 6.873 | 4.038 | 7.05 | 8.121 | 7.365 | 0.394 | 0.365 | 0.476 | 0.292 | 0.494 | 0.575 | 0.487 |
| 5 | 0.17 | 0.546 | 0.532 | 0.429 | 0.525 | 0.561 | 0.308 | 0.552 | 9.012 | 8.547 | 7.104 | 4.863 | 7.329 | 3.588 | 8.214 | 1.53 | 1.42 | 1.208 | 0.824 | 1.239 | 0.602 | 1.385 |
| 6 | 0.47 | 0.022 | 0.03 | 0.028 | 0.133 | 0.039 | 0.035 | 0.047 | 3.261 | 3.345 | 2.241 | 2.928 | 2.541 | 0.834 | 2.682 | 1.515 | 1.603 | 1.058 | 1.375 | 1.159 | 0.396 | 1.278 |
| | | i3+3 | mTPI | mTPI-2 | 3+3 | BOIN | BLRM | CRM | | | | | | | | | | | | | | |
| Prob. of Select MTD | | 0.546 | 0.532 | 0.429 | 0.525 | 0.561 | 0.308 | 0.552 | | | | | | | | | | | | | | |
| Prob. of Toxicity | | 0.126 | 0.126 | 0.106 | 0.129 | 0.11 | 0.074 | 0.118 | | | | | | | | | | | | | | |
| Prob. of Select Dose-over-MTD | | 0.022 | 0.03 | 0.028 | 0.133 | 0.039 | 0.035 | 0.047 | | | | | | | | | | | | | | |
| Prob. of No Selection | | 0 | 0 | 0 | 0.003 | 0 | 0.009 | 0 | | | | | | | | | | | | | | |

## Scenario 5

| Target Toxicity Prob. = 0.17 | | Selection Prob. | | | | | | | # of Patients Treated | | | | | | | # of Toxicities | | | | | | |
|---|---|---|---|---|---|---|---|---|---|---|---|---|---|---|---|---|---|---|---|---|---|---|
| Dose Level | True Tox Prob. | i3+3 | mTPI | mTPI-2 | 3+3 | BOIN | BLRM | CRM | i3+3 | mTPI | mTPI-2 | 3+3 | BOIN | BLRM | CRM | i3+3 | mTPI | mTPI-2 | 3+3 | BOIN | BLRM | CRM |
| 1 | 0.02 | 0.995 | 0.991 | 0.991 | 0.862 | 0.934 | 0.88 | 0.92 | 22.272 | 22.077 | 22.5 | 5.622 | 22.746 | 21.528 | 22.671 | 0.415 | 0.437 | 0.438 | 0.114 | 0.45 | 0.444 | 0.436 |
| 2 | 0.47 | 0.004 | 0.007 | 0.007 | 0.132 | 0.064 | 0.057 | 0.078 | 7.131 | 7.392 | 7.017 | 4.56 | 6.726 | 6.516 | 6.675 | 3.368 | 3.466 | 3.285 | 2.173 | 3.199 | 3.024 | 3.152 |
| 3 | 0.77 | 0 | 0 | 0 | 0.002 | 0 | 0 | 0 | 0.585 | 0.471 | 0.423 | 0.705 | 0.468 | 0.444 | 0.594 | 0.449 | 0.365 | 0.326 | 0.539 | 0.358 | 0.339 | 0.461 |
| 4 | 0.87 | 0 | 0 | 0 | 0 | 0 | 0 | 0 | 0.012 | 0.006 | 0.006 | 0.018 | 0.006 | 0 | 0.006 | 0.012 | 0.005 | 0.005 | 0.015 | 0.004 | 0 | 0.005 |
| 5 | 0.92 | 0 | 0 | 0 | 0 | 0 | 0 | 0 | 0 | 0 | 0 | 0 | 0 | 0 | 0 | 0 | 0 | 0 | 0 | 0 | 0 | 0 |
| 6 | 0.96 | 0 | 0 | 0 | 0 | 0 | 0 | 0 | 0 | 0 | 0 | 0 | 0 | 0 | 0 | 0 | 0 | 0 | 0 | 0 | 0 | 0 |
| | | i3+3 | mTPI | mTPI-2 | 3+3 | BOIN | BLRM | CRM | | | | | | | | | | | | | | |
| Prob. of Select MTD | | 0.995 | 0.991 | 0.991 | 0.862 | 0.934 | 0.88 | 0.92 | | | | | | | | | | | | | | |
| Prob. of Toxicity | | 0.141 | 0.143 | 0.135 | 0.261 | 0.134 | 0.134 | 0.135 | | | | | | | | | | | | | | |
| Prob. of Select Dose-over-MTD | | 0.004 | 0.007 | 0.007 | 0.134 | 0.064 | 0.057 | 0.078 | | | | | | | | | | | | | | |
| Prob. of No Selection | | 0.001 | 0.002 | 0.002 | 0.004 | 0.002 | 0.063 | 0.002 | | | | | | | | | | | | | | |

### Scenario 6

| Target Toxicity Prob. = 0.17 | | Selection Prob. | | | | | | # of Patients Treated | | | | | | # of Toxicities | | | | | |
|---|---|---|---|---|---|---|---|---|---|---|---|---|---|---|---|---|---|---|---|---|
| Dose Level | True Tox Prob. | i3+3 | mTPI | mTPI-2 | 3+3 | BOIN | BLRM | CRM | i3+3 | mTPI | mTPI-2 | 3+3 | BOIN | BLRM | CRM | i3+3 | mTPI | mTPI-2 | 3+3 | BOIN | BLRM | CRM |
| 1 | 0.01 | 0 | 0.001 | 0.001 | 0.009 | 0.002 | 0.001 | 0.001 | 3.003 | 3.33 | 3.51 | 3.12 | 3.27 | 3.3 | 3.252 | 0 | 0.034 | 0.035 | 0.033 | 0 | 0.036 | 0.034 |
| 2 | 0.02 | 0.035 | 0.043 | 0.077 | 0.057 | 0.032 | 0.05 | 0.045 | 3.735 | 3.999 | 5.139 | 3.357 | 4.881 | 5.388 | 4.62 | 0.053 | 0.074 | 0.106 | 0.08 | 0.095 | 0.115 | 0.095 |
| 3 | 0.07 | 0.945 | 0.932 | 0.895 | 0.806 | 0.898 | 0.861 | 0.877 | 16.353 | 15.771 | 15.831 | 5.571 | 15.75 | 16.797 | 15.78 | 1.108 | 1.114 | 1.097 | 0.391 | 1.077 | 1.161 | 1.106 |
| 4 | 0.47 | 0.02 | 0.024 | 0.025 | 0.122 | 0.066 | 0.078 | 0.077 | 6.273 | 6.366 | 5.1 | 4.317 | 5.592 | 4.131 | 5.871 | 2.915 | 2.937 | 2.413 | 2.062 | 2.586 | 1.964 | 2.696 |
| 5 | 0.67 | 0 | 0 | 0.002 | 0.004 | 0.002 | 0.001 | 0 | 0.612 | 0.519 | 0.405 | 0.732 | 0.48 | 0.141 | 0.468 | 0.404 | 0.362 | 0.279 | 0.483 | 0.319 | 0.097 | 0.326 |
| 6 | 0.87 | 0 | 0 | 0 | 0 | 0 | 0 | 0 | 0.024 | 0.015 | 0.015 | 0.018 | 0.027 | 0 | 0.009 | 0.021 | 0.013 | 0.013 | 0.015 | 0.022 | 0 | 0.008 |
| | | i3+3 | mTPI | mTPI-2 | 3+3 | BOIN | BLRM | CRM | | | | | | | | | | | | | | |
| Prob. of Select MTD | | 0.945 | 0.932 | 0.895 | 0.806 | 0.898 | 0.861 | 0.877 | | | | | | | | | | | | | | |
| Prob. of Toxicity | | 0.15 | 0.151 | 0.131 | 0.179 | 0.137 | 0.113 | 0.142 | | | | | | | | | | | | | | |
| Prob. of Select Dose-over-MTD | | 0.02 | 0.024 | 0.027 | 0.126 | 0.068 | 0.079 | 0.077 | | | | | | | | | | | | | | |
| Prob. of No Selection | | 0 | 0 | 0 | 0.002 | 0 | 0.009 | 0 | | | | | | | | | | | | | | |

### Scenario 7

| Target Toxicity Prob. = 0.17 | | Selection Prob. | | | | | | # of Patients Treated | | | | | | | # of Toxicities | | | | | | |
|---|---|---|---|---|---|---|---|---|---|---|---|---|---|---|---|---|---|---|---|---|---|
| Dose Level | True Tox Prob. | i3+3 | mTPI | mTPI-2 | 3+3 | BOIN | BLRM | CRM | i3+3 | mTPI | mTPI-2 | 3+3 | BOIN | BLRM | CRM | i3+3 | mTPI | mTPI-2 | 3+3 | BOIN | BLRM | CRM |
| 1 | 0.01 | 0 | 0 | 0 | 0.001 | 0 | 0 | 0 | 3 | 3.306 | 3.402 | 3.084 | 3 | 3.249 | 3.21 | 0 | 0.034 | 0.034 | 0.03 | 0 | 0.036 | 0.034 |
| 2 | 0.01 | 0.002 | 0.008 | 0.014 | 0.017 | 0.006 | 0.009 | 0.011 | 3.078 | 3.435 | 3.975 | 3.12 | 3.561 | 4.212 | 3.396 | 0 | 0.032 | 0.04 | 0.029 | 0 | 0.042 | 0.032 |
| 3 | 0.04 | 0.035 | 0.043 | 0.099 | 0.064 | 0.048 | 0.126 | 0.042 | 4.143 | 4.731 | 5.397 | 3.504 | 5.097 | 8.607 | 4.338 | 0.132 | 0.214 | 0.232 | 0.145 | 0.203 | 0.345 | 0.191 |
| 4 | 0.07 | 0.224 | 0.235 | 0.268 | 0.14 | 0.206 | 0.479 | 0.275 | 5.679 | 5.436 | 6.231 | 3.834 | 6.354 | 8.793 | 7.41 | 0.384 | 0.344 | 0.433 | 0.3 | 0.456 | 0.632 | 0.498 |
| 5 | 0.12 | 0.738 | 0.71 | 0.61 | 0.746 | 0.738 | 0.356 | 0.651 | 11.223 | 10.434 | 8.661 | 5.322 | 9.498 | 3.879 | 9.279 | 1.361 | 1.28 | 1.052 | 0.642 | 1.072 | 0.455 | 1.126 |
| 6 | 0.67 | 0.001 | 0.004 | 0.009 | 0.029 | 0.002 | 0.021 | 0.021 | 2.877 | 2.658 | 2.334 | 2.934 | 2.49 | 1.017 | 2.367 | 1.894 | 1.803 | 1.562 | 1.972 | 1.675 | 0.685 | 1.604 |
| | | i3+3 | mTPI | mTPI-2 | 3+3 | BOIN | BLRM | CRM | | | | | | | | | | | | | | |
| Prob. of Select MTD | | 0.738 | 0.71 | 0.61 | 0.746 | 0.738 | 0.356 | 0.651 | | | | | | | | | | | | | | |
| Prob. of Toxicity | | 0.126 | 0.124 | 0.112 | 0.143 | 0.112 | 0.074 | 0.116 | | | | | | | | | | | | | | |
| Prob. of Select Dose-over-MTD | | 0.001 | 0.004 | 0.009 | 0.029 | 0.002 | 0.021 | 0.021 | | | | | | | | | | | | | | |
| Prob. of No Selection | | 0 | 0 | 0 | 0.003 | 0 | 0.009 | 0 | | | | | | | | | | | | | | |

### Scenario 8

| Target Toxicity Prob. = 0.17 | | Selection Prob. | | | | | | # of Patients Treated | | | | | | | # of Toxicities | | | | | | |
|---|---|---|---|---|---|---|---|---|---|---|---|---|---|---|---|---|---|---|---|---|---|
| Dose Level | True Tox Prob. | i3+3 | mTPI | mTPI-2 | 3+3 | BOIN | BLRM | CRM | i3+3 | mTPI | mTPI-2 | 3+3 | BOIN | BLRM | CRM | i3+3 | mTPI | mTPI-2 | 3+3 | BOIN | BLRM | CRM |
| 1 | 0.16 | 0.291 | 0.325 | 0.399 | 0.203 | 0.317 | 0.248 | 0.31 | 11.928 | 11.901 | 15.285 | 4.497 | 13.077 | 9.639 | 12.231 | 1.938 | 1.92 | 2.449 | 0.731 | 2.061 | 1.495 | 1.965 |
| 2 | 0.18 | 0.198 | 0.193 | 0.185 | 0.174 | 0.229 | 0.251 | 0.219 | 6.762 | 7.458 | 6.189 | 3.603 | 6.78 | 7.971 | 6.516 | 1.186 | 1.333 | 1.127 | 0.655 | 1.205 | 1.456 | 1.173 |
| 3 | 0.2 | 0.143 | 0.156 | 0.107 | 0.146 | 0.15 | 0.148 | 0.183 | 4.125 | 4.194 | 3.234 | 2.754 | 3.753 | 4.398 | 4.758 | 0.828 | 0.819 | 0.628 | 0.556 | 0.72 | 0.848 | 0.901 |
| 4 | 0.22 | 0.08 | 0.088 | 0.058 | 0.085 | 0.077 | 0.051 | 0.101 | 2.364 | 2.22 | 1.449 | 1.884 | 1.869 | 1.101 | 2.436 | 0.537 | 0.492 | 0.313 | 0.439 | 0.412 | 0.251 | 0.539 |
| 5 | 0.24 | 0.06 | 0.04 | 0.024 | 0.076 | 0.052 | 0.01 | 0.039 | 1.047 | 0.891 | 0.63 | 1.272 | 0.9 | 0.231 | 0.933 | 0.259 | 0.213 | 0.149 | 0.291 | 0.197 | 0.047 | 0.227 |
| 6 | 0.26 | 0.021 | 0.015 | 0.007 | 0.089 | 0.024 | 0.005 | 0.014 | 0.579 | 0.429 | 0.273 | 0.759 | 0.495 | 0.057 | 0.339 | 0.157 | 0.123 | 0.081 | 0.211 | 0.128 | 0.008 | 0.1 |
| | | i3+3 | mTPI | mTPI-2 | 3+3 | BOIN | BLRM | CRM | | | | | | | | | | | | | | |
| Prob. of Select MTD | | 0.712 | 0.762 | 0.749 | 0.608 | 0.773 | 0.698 | 0.813 | | | | | | | | | | | | | | |
| Prob. of Toxicity | | 0.183 | 0.181 | 0.175 | 0.195 | 0.176 | 0.175 | 0.18 | | | | | | | | | | | | | | |
| Prob. of Select Dose-over-MTD | | 0.081 | 0.055 | 0.031 | 0.165 | 0.076 | 0.015 | 0.053 | | | | | | | | | | | | | | |
| Prob. of No Selection | | 0.207 | 0.183 | 0.22 | 0.227 | 0.151 | 0.287 | 0.134 | | | | | | | | | | | | | | |

### Scenario 9

| Target Toxicity Prob. = 0.17 | | Selection Prob. | | | | | | # of Patients Treated | | | | | | | # of Toxicities | | | | | | |
|---|---|---|---|---|---|---|---|---|---|---|---|---|---|---|---|---|---|---|---|---|---|
| Dose Level | True Tox Prob. | i3+3 | mTPI | mTPI-2 | 3+3 | BOIN | BLRM | CRM | i3+3 | mTPI | mTPI-2 | 3+3 | BOIN | BLRM | CRM | i3+3 | mTPI | mTPI-2 | 3+3 | BOIN | BLRM | CRM |
| 1 | 0.12 | 0.196 | 0.25 | 0.315 | 0.142 | 0.224 | 0.155 | 0.19 | 9.285 | 10.005 | 12.327 | 4.191 | 10.686 | 7.749 | 9.531 | 1.164 | 1.189 | 1.472 | 0.519 | 1.253 | 0.907 | 1.152 |
| 2 | 0.14 | 0.19 | 0.201 | 0.224 | 0.148 | 0.224 | 0.266 | 0.234 | 6.834 | 7.68 | 7.302 | 3.726 | 7.122 | 8.52 | 7.239 | 0.9 | 1.079 | 1.053 | 0.513 | 1.011 | 1.182 | 1.008 |
| 3 | 0.16 | 0.187 | 0.201 | 0.167 | 0.15 | 0.196 | 0.232 | 0.209 | 5.295 | 5.208 | 4.311 | 3.189 | 4.983 | 6.405 | 5.589 | 0.815 | 0.846 | 0.684 | 0.508 | 0.758 | 1.009 | 0.921 |
| 4 | 0.18 | 0.149 | 0.145 | 0.095 | 0.122 | 0.117 | 0.124 | 0.165 | 3.582 | 3.165 | 2.496 | 2.568 | 2.913 | 2.301 | 3.576 | 0.64 | 0.577 | 0.44 | 0.48 | 0.516 | 0.41 | 0.62 |
| 5 | 0.2 | 0.115 | 0.063 | 0.067 | 0.104 | 0.103 | 0.03 | 0.083 | 2.043 | 1.623 | 1.296 | 2.007 | 1.569 | 0.441 | 1.686 | 0.408 | 0.334 | 0.252 | 0.413 | 0.283 | 0.1 | 0.349 |
| 6 | 0.22 | 0.067 | 0.064 | 0.038 | 0.185 | 0.064 | 0.005 | 0.047 | 1.488 | 1.026 | 0.753 | 1.278 | 1.002 | 0.09 | 0.804 | 0.351 | 0.221 | 0.171 | 0.304 | 0.213 | 0.026 | 0.174 |
| | | i3+3 | mTPI | mTPI-2 | 3+3 | BOIN | BLRM | CRM | | | | | | | | | | | | | | |
| Prob. of Select MTD | | 0.904 | 0.924 | 0.906 | 0.851 | 0.928 | 0.812 | 0.928 | | | | | | | | | | | | | | |
| Prob. of Toxicity | | 0.15 | 0.148 | 0.143 | 0.161 | 0.143 | 0.142 | 0.149 | | | | | | | | | | | | | | |
| Prob. of Select Dose-over-MTD | | 0 | 0 | 0 | 0 | 0 | 0 | 0 | | | | | | | | | | | | | | |
| Prob. of No Selection | | 0.096 | 0.076 | 0.094 | 0.149 | 0.072 | 0.188 | 0.072 | | | | | | | | | | | | | | |

### Scenario 10

| Target Toxicity Prob. = 0.17 | | Selection Prob. | | | | | | # of Patients Treated | | | | | | | # of Toxicities | | | | | | |
|---|---|---|---|---|---|---|---|---|---|---|---|---|---|---|---|---|---|---|---|---|---|
| Dose Level | True Tox Prob. | i3+3 | mTPI | mTPI-2 | 3+3 | BOIN | BLRM | CRM | i3+3 | mTPI | mTPI-2 | 3+3 | BOIN | BLRM | CRM | i3+3 | mTPI | mTPI-2 | 3+3 | BOIN | BLRM | CRM |
| 1 | 0.08 | 0.112 | 0.125 | 0.179 | 0.075 | 0.082 | 0.054 | 0.083 | 6.78 | 7.362 | 9.39 | 3.834 | 7.404 | 5.649 | 6.927 | 0.528 | 0.589 | 0.735 | 0.336 | 0.568 | 0.443 | 0.559 |
| 2 | 0.1 | 0.157 | 0.171 | 0.226 | 0.125 | 0.171 | 0.213 | 0.184 | 6.759 | 7.266 | 7.542 | 3.765 | 6.984 | 8.229 | 6.663 | 0.703 | 0.723 | 0.787 | 0.352 | 0.672 | 0.765 | 0.672 |
| 3 | 0.12 | 0.166 | 0.202 | 0.224 | 0.125 | 0.213 | 0.298 | 0.236 | 5.454 | 5.76 | 5.4 | 3.585 | 5.718 | 8.265 | 6.639 | 0.642 | 0.7 | 0.635 | 0.463 | 0.667 | 0.971 | 0.798 |
| 4 | 0.14 | 0.182 | 0.221 | 0.155 | 0.126 | 0.179 | 0.209 | 0.208 | 4.38 | 4.236 | 3.522 | 3.114 | 4.218 | 3.966 | 4.722 | 0.609 | 0.6 | 0.539 | 0.45 | 0.594 | 0.543 | 0.665 |
| 5 | 0.16 | 0.204 | 0.101 | 0.086 | 0.118 | 0.164 | 0.095 | 0.145 | 3.192 | 2.463 | 1.974 | 2.745 | 2.793 | 1.245 | 2.577 | 0.521 | 0.395 | 0.314 | 0.442 | 0.448 | 0.209 | 0.425 |
| 6 | 0.18 | 0.158 | 0.162 | 0.111 | 0.355 | 0.16 | 0.037 | 0.126 | 2.892 | 2.442 | 1.701 | 1.932 | 2.226 | 0.348 | 2.001 | 0.523 | 0.445 | 0.305 | 0.346 | 0.399 | 0.067 | 0.356 |
| | | i3+3 | mTPI | mTPI-2 | 3+3 | BOIN | BLRM | CRM | | | | | | | | | | | | | | |
| Prob. of Select MTD | | 0.71 | 0.686 | 0.576 | 0.724 | 0.72 | 0.639 | 0.715 | | | | | | | | | | | | | | |
| Prob. of Toxicity | | 0.12 | 0.117 | 0.112 | 0.126 | 0.114 | 0.108 | 0.118 | | | | | | | | | | | | | | |
| Prob. of Select Dose-over-MTD | | 0 | 0 | 0 | 0 | 0 | 0 | 0 | | | | | | | | | | | | | | |
| Prob. of No Selection | | 0.021 | 0.018 | 0.019 | 0.076 | 0.027 | 0.094 | 0.018 | | | | | | | | | | | | | | |

## Scenario 11

| Target Toxicity Prob. = 0.17 | | | | Selection Prob. | | | | | | # of Patients Treated | | | | | | | # of Toxicities | | | | |
|---|---|---|---|---|---|---|---|---|---|---|---|---|---|---|---|---|---|---|---|---|---|
| Dose Level | True Tox Prob. | i3+3 | mTPI | mTPI-2 | 3+3 | BOIN | BLRM | CRM | i3+3 | mTPI | mTPI-2 | 3+3 | BOIN | BLRM | CRM | i3+3 | mTPI | mTPI-2 | 3+3 | BOIN | BLRM | CRM |
| 1 | 0.02 | 0.034 | 0.036 | 0.066 | 0.06 | 0.034 | 0.027 | 0.031 | 4.017 | 4.107 | 5.454 | 3.309 | 5.409 | 4.374 | 4.167 | 0.067 | 0.071 | 0.106 | 0.056 | 0.103 | 0.093 | 0.072 |
| 2 | 0.08 | 0.195 | 0.188 | 0.317 | 0.187 | 0.193 | 0.263 | 0.142 | 7.059 | 7.554 | 9.138 | 4.104 | 8.427 | 9.513 | 6.558 | 0.531 | 0.609 | 0.749 | 0.348 | 0.695 | 0.733 | 0.551 |
| 3 | 0.14 | 0.335 | 0.329 | 0.323 | 0.22 | 0.347 | 0.412 | 0.346 | 8.313 | 7.863 | 7.866 | 4.23 | 7.824 | 10.479 | 8.601 | 1.128 | 1.12 | 1.127 | 0.627 | 1.042 | 1.436 | 1.166 |
| 4 | 0.2 | 0.262 | 0.304 | 0.196 | 0.238 | 0.247 | 0.223 | 0.319 | 6.081 | 6.27 | 4.863 | 3.648 | 5.028 | 4.32 | 6.819 | 1.246 | 1.211 | 0.948 | 0.716 | 0.979 | 0.789 | 1.3 |
| 5 | 0.26 | 0.137 | 0.107 | 0.076 | 0.155 | 0.141 | 0.056 | 0.126 | 3.144 | 3.087 | 1.995 | 2.715 | 2.322 | 0.771 | 2.979 | 0.819 | 0.811 | 0.516 | 0.721 | 0.577 | 0.194 | 0.798 |
| 6 | 0.32 | 0.036 | 0.034 | 0.02 | 0.136 | 0.036 | 0.005 | 0.034 | 1.359 | 1.065 | 0.63 | 1.305 | 0.936 | 0.171 | 0.822 | 0.435 | 0.339 | 0.197 | 0.437 | 0.303 | 0.054 | 0.259 |
| | | i3+3 | mTPI | mTPI-2 | 3+3 | BOIN | BLRM | CRM | | | | | | | | | | | | | | |
| Prob. of Select MTD | | 0.597 | 0.633 | 0.519 | 0.458 | 0.594 | 0.635 | 0.665 | | | | | | | | | | | | | | |
| Prob. of Toxicity | | 0.141 | 0.139 | 0.122 | 0.15 | 0.124 | 0.111 | 0.138 | | | | | | | | | | | | | | |
| Prob. of Select Dose-over-MTD | | 0.173 | 0.141 | 0.096 | 0.291 | 0.177 | 0.061 | 0.16 | | | | | | | | | | | | | | |
| Prob. of No Selection | | 0.001 | 0.002 | 0.002 | 0.004 | 0.002 | 0.014 | 0.002 | | | | | | | | | | | | | | |

## Scenario 12

| Target Toxicity Prob. = 0.17 | | | | Selection Prob. | | | | | | # of Patients Treated | | | | | | | # of Toxicities | | | | |
|---|---|---|---|---|---|---|---|---|---|---|---|---|---|---|---|---|---|---|---|---|---|
| Dose Level | True Tox Prob. | i3+3 | mTPI | mTPI-2 | 3+3 | BOIN | BLRM | CRM | i3+3 | mTPI | mTPI-2 | 3+3 | BOIN | BLRM | CRM | i3+3 | mTPI | mTPI-2 | 3+3 | BOIN | BLRM | CRM |
| 1 | 0.02 | 0.027 | 0.027 | 0.049 | 0.046 | 0.026 | 0.023 | 0.024 | 3.969 | 3.969 | 5.127 | 3.279 | 5.043 | 4.215 | 3.969 | 0.078 | 0.07 | 0.102 | 0.063 | 0.097 | 0.09 | 0.071 |
| 2 | 0.07 | 0.151 | 0.138 | 0.258 | 0.157 | 0.154 | 0.188 | 0.103 | 6.273 | 6.87 | 8.196 | 3.96 | 7.71 | 8.325 | 5.907 | 0.429 | 0.498 | 0.585 | 0.286 | 0.554 | 0.559 | 0.446 |
| 3 | 0.12 | 0.27 | 0.282 | 0.295 | 0.188 | 0.304 | 0.394 | 0.294 | 7.443 | 7.14 | 7.56 | 4.143 | 7.47 | 10.521 | 7.89 | 0.915 | 0.877 | 0.941 | 0.56 | 0.888 | 1.223 | 0.922 |
| 4 | 0.17 | 0.357 | 0.389 | 0.286 | 0.281 | 0.308 | 0.303 | 0.376 | 6.756 | 6.897 | 5.856 | 3.834 | 5.757 | 5.289 | 7.677 | 1.114 | 1.123 | 0.955 | 0.649 | 0.943 | 0.838 | 1.244 |
| 5 | 0.27 | 0.152 | 0.126 | 0.086 | 0.187 | 0.171 | 0.069 | 0.172 | 3.87 | 3.882 | 2.514 | 3.105 | 2.901 | 1.098 | 3.657 | 1.063 | 1.057 | 0.686 | 0.848 | 0.754 | 0.297 | 1.014 |
| 6 | 0.34 | 0.042 | 0.036 | 0.024 | 0.134 | 0.035 | 0.009 | 0.029 | 1.662 | 1.188 | 0.693 | 1.425 | 1.065 | 0.18 | 0.846 | 0.554 | 0.41 | 0.237 | 0.51 | 0.353 | 0.058 | 0.292 |
| | | i3+3 | mTPI | mTPI-2 | 3+3 | BOIN | BLRM | CRM | | | | | | | | | | | | | | |
| Prob. of Select MTD | | 0.627 | 0.671 | 0.581 | 0.469 | 0.612 | 0.697 | 0.67 | | | | | | | | | | | | | | |
| Prob. of Toxicity | | 0.139 | 0.135 | 0.117 | 0.148 | 0.12 | 0.103 | 0.133 | | | | | | | | | | | | | | |
| Prob. of Select Dose-over-MTD | | 0.194 | 0.162 | 0.11 | 0.321 | 0.206 | 0.078 | 0.201 | | | | | | | | | | | | | | |
| Prob. of No Selection | | 0.001 | 0.002 | 0.002 | 0.007 | 0.002 | 0.014 | 0.002 | | | | | | | | | | | | | | |

## Scenario 13

| Target Toxicity Prob. = 0.17 | | | | Selection Prob. | | | | | | # of Patients Treated | | | | | | | # of Toxicities | | | | |
|---|---|---|---|---|---|---|---|---|---|---|---|---|---|---|---|---|---|---|---|---|---|
| Dose Level | True Tox Prob. | i3+3 | mTPI | mTPI-2 | 3+3 | BOIN | BLRM | CRM | i3+3 | mTPI | mTPI-2 | 3+3 | BOIN | BLRM | CRM | i3+3 | mTPI | mTPI-2 | 3+3 | BOIN | BLRM | CRM |
| 1 | 0.17 | 0.404 | 0.404 | 0.47 | 0.266 | 0.441 | 0.298 | 0.394 | 13.401 | 12.771 | 16.626 | 4.686 | 15.072 | 10.566 | 13.311 | 2.215 | 2.134 | 2.801 | 0.8 | 2.496 | 1.718 | 2.243 |
| 2 | 0.22 | 0.223 | 0.229 | 0.178 | 0.231 | 0.236 | 0.252 | 0.277 | 7.683 | 8.265 | 6.288 | 3.792 | 7.11 | 7.908 | 7.65 | 1.714 | 1.814 | 1.387 | 0.851 | 1.561 | 1.688 | 1.68 |
| 3 | 0.27 | 0.099 | 0.105 | 0.061 | 0.145 | 0.116 | 0.094 | 0.117 | 3.855 | 3.663 | 2.499 | 2.505 | 3.054 | 3.27 | 3.882 | 1.031 | 1.004 | 0.667 | 0.695 | 0.822 | 0.837 | 1.029 |
| 4 | 0.32 | 0.039 | 0.029 | 0.015 | 0.073 | 0.038 | 0.008 | 0.032 | 1.419 | 1.278 | 0.735 | 1.365 | 1.101 | 0.477 | 1.317 | 0.471 | 0.416 | 0.245 | 0.44 | 0.345 | 0.151 | 0.403 |
| 5 | 0.37 | 0.008 | 0.008 | 0.002 | 0.025 | 0.004 | 0.002 | 0.008 | 0.432 | 0.321 | 0.174 | 0.639 | 0.285 | 0.027 | 0.267 | 0.158 | 0.112 | 0.059 | 0.242 | 0.113 | 0.008 | 0.099 |
| 6 | 0.42 | 0 | 0 | 0 | 0.013 | 0 | 0 | 0 | 0.078 | 0.057 | 0.033 | 0.165 | 0.042 | 0.012 | 0.048 | 0.04 | 0.028 | 0.014 | 0.07 | 0.015 | 0.005 | 0.024 |
| | | i3+3 | mTPI | mTPI-2 | 3+3 | BOIN | BLRM | CRM | | | | | | | | | | | | | | |
| Prob. of Select MTD | | 0.627 | 0.633 | 0.648 | 0.497 | 0.677 | 0.55 | 0.671 | | | | | | | | | | | | | | |
| Prob. of Toxicity | | 0.21 | 0.209 | 0.196 | 0.236 | 0.2 | 0.198 | 0.207 | | | | | | | | | | | | | | |
| Prob. of Select Dose-over-MTD | | 0.146 | 0.142 | 0.078 | 0.256 | 0.158 | 0.104 | 0.157 | | | | | | | | | | | | | | |
| Prob. of No Selection | | 0.227 | 0.225 | 0.274 | 0.247 | 0.165 | 0.346 | 0.172 | | | | | | | | | | | | | | |

## Scenario 14

| Target Toxicity Prob. = 0.17 | | | | Selection Prob. | | | | | | # of Patients Treated | | | | | | | # of Toxicities | | | | |
|---|---|---|---|---|---|---|---|---|---|---|---|---|---|---|---|---|---|---|---|---|---|
| Dose Level | True Tox Prob. | i3+3 | mTPI | mTPI-2 | 3+3 | BOIN | BLRM | CRM | i3+3 | mTPI | mTPI-2 | 3+3 | BOIN | BLRM | CRM | i3+3 | mTPI | mTPI-2 | 3+3 | BOIN | BLRM | CRM |
| 1 | 0.02 | 0.01 | 0.012 | 0.015 | 0.024 | 0.014 | 0.008 | 0.011 | 3.744 | 3.72 | 4.416 | 3.207 | 4.488 | 3.843 | 3.636 | 0.083 | 0.066 | 0.084 | 0.058 | 0.087 | 0.085 | 0.066 |
| 2 | 0.05 | 0.053 | 0.044 | 0.11 | 0.06 | 0.074 | 0.081 | 0.036 | 4.653 | 5.22 | 6.111 | 3.54 | 5.952 | 6.405 | 4.464 | 0.214 | 0.275 | 0.311 | 0.18 | 0.324 | 0.308 | 0.238 |
| 3 | 0.08 | 0.119 | 0.147 | 0.204 | 0.109 | 0.157 | 0.256 | 0.143 | 5.691 | 5.784 | 6.39 | 3.876 | 6.135 | 9.354 | 6.183 | 0.441 | 0.428 | 0.498 | 0.326 | 0.482 | 0.674 | 0.457 |
| 4 | 0.11 | 0.2 | 0.275 | 0.257 | 0.128 | 0.241 | 0.338 | 0.277 | 5.616 | 5.76 | 5.637 | 3.798 | 5.454 | 6.633 | 6.618 | 0.618 | 0.614 | 0.615 | 0.441 | 0.579 | 0.677 | 0.709 |
| 5 | 0.14 | 0.288 | 0.199 | 0.198 | 0.153 | 0.23 | 0.199 | 0.256 | 4.749 | 4.68 | 4.101 | 3.504 | 3.954 | 2.499 | 4.881 | 0.624 | 0.696 | 0.601 | 0.475 | 0.536 | 0.347 | 0.7 |
| 6 | 0.17 | 0.33 | 0.321 | 0.214 | 0.518 | 0.282 | 0.104 | 0.275 | 5.547 | 4.782 | 3.291 | 2.709 | 3.963 | 0.894 | 4.164 | 0.954 | 0.809 | 0.567 | 0.473 | 0.638 | 0.146 | 0.719 |
| | | i3+3 | mTPI | mTPI-2 | 3+3 | BOIN | BLRM | CRM | | | | | | | | | | | | | | |
| Prob. of Select MTD | | 0.618 | 0.52 | 0.412 | 0.671 | 0.512 | 0.303 | 0.531 | | | | | | | | | | | | | | |
| Prob. of Toxicity | | 0.098 | 0.096 | 0.089 | 0.095 | 0.088 | 0.076 | 0.096 | | | | | | | | | | | | | | |
| Prob. of Select Dose-over-MTD | | 0 | 0 | 0 | 0 | 0 | 0 | 0 | | | | | | | | | | | | | | |
| Prob. of No Selection | | 0 | 0.002 | 0.002 | 0.008 | 0.002 | 0.014 | 0.002 | | | | | | | | | | | | | | |

## Scenario 1

Target Toxicity Prob. = 0.3

| Dose Level | True Tox Prob. | Selection Prob. | | | | | | | # of Patients Treated | | | | | | | # of Toxicities | | | | | | |
|---|---|---|---|---|---|---|---|---|---|---|---|---|---|---|---|---|---|---|---|---|---|---|
| | | i3+3 | mTPI | mTPI-2 | 3+3 | BOIN | BLRM | CRM | i3+3 | mTPI | mTPI-2 | 3+3 | BOIN | BLRM | CRM | i3+3 | mTPI | mTPI-2 | 3+3 | BOIN | BLRM | CRM |
| 1 | 0.02 | 0 | 0.001 | 0.001 | 0.02 | 0 | 0.002 | 0.001 | 3.249 | 3.255 | 3.246 | 3.21 | 3.219 | 3.189 | 3.228 | 0.077 | 0.069 | 0.069 | 0.064 | 0.062 | 0.069 | 0.069 |
| 2 | 0.05 | 0.005 | 0.002 | 0.004 | 0.108 | 0.004 | 0.002 | 0.002 | 3.66 | 3.747 | 3.708 | 3.651 | 3.813 | 3.678 | 3.114 | 0.176 | 0.181 | 0.18 | 0.183 | 0.195 | 0.166 | 0.147 |
| 3 | 0.1 | 0.03 | 0.051 | 0.04 | 0.165 | 0.026 | 0.066 | 0.016 | 4.74 | 5.154 | 4.866 | 4.044 | 4.893 | 6.141 | 3.711 | 0.479 | 0.543 | 0.512 | 0.417 | 0.503 | 0.586 | 0.4 |
| 4 | 0.15 | 0.134 | 0.178 | 0.145 | 0.228 | 0.126 | 0.204 | 0.089 | 5.898 | 5.886 | 5.676 | 3.921 | 5.901 | 7.059 | 5.037 | 0.846 | 0.841 | 0.808 | 0.586 | 0.895 | 1.053 | 0.744 |
| 5 | 0.2 | 0.279 | 0.273 | 0.272 | 0.174 | 0.258 | 0.296 | 0.269 | 5.694 | 5.724 | 5.832 | 3.375 | 5.616 | 5.289 | 6.15 | 1.131 | 1.171 | 1.186 | 0.711 | 1.099 | 1.045 | 1.209 |
| 6 | 0.25 | 0.552 | 0.495 | 0.538 | 0.297 | 0.586 | 0.43 | 0.623 | 6.759 | 6.234 | 6.672 | 2.028 | 6.558 | 4.644 | 8.76 | 1.684 | 1.533 | 1.64 | 0.499 | 1.59 | 1.149 | 2.17 |

| | i3+3 | mTPI | mTPI-2 | 3+3 | BOIN | BLRM | CRM |
|---|---|---|---|---|---|---|---|
| Prob. of Select MTD | 0.552 | 0.495 | 0.538 | 0.297 | 0.586 | 0.43 | 0.623 |
| Prob. of Toxicity | 0.146 | 0.145 | 0.147 | 0.122 | 0.145 | 0.136 | 0.158 |
| Prob. of Select Dose-over-MTD | 0 | 0 | 0 | 0 | 0 | 0 | 0 |
| Prob. of No Selection | 0 | 0 | 0 | 0.008 | 0 | 0 | 0 |

## Scenario 2

Target Toxicity Prob. = 0.3

| Dose Level | True Tox Prob. | Selection Prob. | | | | | | | # of Patients Treated | | | | | | | # of Toxicities | | | | | | |
|---|---|---|---|---|---|---|---|---|---|---|---|---|---|---|---|---|---|---|---|---|---|---|
| | | i3+3 | mTPI | mTPI-2 | 3+3 | BOIN | BLRM | CRM | i3+3 | mTPI | mTPI-2 | 3+3 | BOIN | BLRM | CRM | i3+3 | mTPI | mTPI-2 | 3+3 | BOIN | BLRM | CRM |
| 1 | 0.35 | 0.432 | 0.45 | 0.436 | 0.265 | 0.533 | 0.292 | 0.539 | 17.847 | 17.625 | 17.964 | 5.046 | 17.607 | 11.922 | 17.415 | 6.228 | 6.134 | 6.249 | 1.73 | 6.143 | 4.248 | 6.041 |
| 2 | 0.45 | 0.086 | 0.087 | 0.087 | 0.057 | 0.11 | 0.055 | 0.118 | 5.19 | 5.607 | 5.172 | 1.875 | 5.067 | 4.203 | 5.004 | 2.313 | 2.495 | 2.306 | 0.847 | 2.265 | 1.975 | 2.236 |
| 3 | 0.5 | 0.018 | 0.013 | 0.02 | 0.01 | 0.026 | 0.011 | 0.027 | 1.08 | 0.933 | 1.011 | 0.459 | 1.212 | 0.72 | 1.698 | 0.554 | 0.476 | 0.517 | 0.233 | 0.592 | 0.377 | 0.844 |
| 4 | 0.6 | 0.001 | 0 | 0.001 | 0 | 0.004 | 0 | 0 | 0.15 | 0.114 | 0.138 | 0.069 | 0.21 | 0.087 | 0.309 | 0.092 | 0.075 | 0.088 | 0.046 | 0.116 | 0.048 | 0.195 |
| 5 | 0.7 | 0 | 0 | 0 | 0 | 0 | 0 | 0 | 0.006 | 0 | 0.003 | 0.009 | 0.015 | 0 | 0.027 | 0.005 | 0 | 0.003 | 0.005 | 0.012 | 0 | 0.019 |
| 6 | 0.8 | 0 | 0 | 0 | 0 | 0 | 0 | 0 | 0 | 0 | 0 | 0 | 0 | 0 | 0 | 0 | 0 | 0 | 0 | 0 | 0 | 0 |

| | i3+3 | mTPI | mTPI-2 | 3+3 | BOIN | BLRM | CRM |
|---|---|---|---|---|---|---|---|
| Prob. of Select MTD | 0.432 | 0.45 | 0.436 | 0.265 | 0.533 | 0.292 | 0.539 |
| Prob. of Toxicity | 0.379 | 0.378 | 0.377 | 0.384 | 0.379 | 0.393 | 0.382 |
| Prob. of Select Dose-over-MTD | 0.105 | 0.1 | 0.108 | 0.067 | 0.14 | 0.066 | 0.145 |
| Prob. of No Selection | 0.463 | 0.45 | 0.456 | 0.668 | 0.327 | 0.642 | 0.316 |

## Scenario 3

Target Toxicity Prob. = 0.3

| Dose Level | True Tox Prob. | Selection Prob. | | | | | | | # of Patients Treated | | | | | | | # of Toxicities | | | | | | |
|---|---|---|---|---|---|---|---|---|---|---|---|---|---|---|---|---|---|---|---|---|---|---|
| | | i3+3 | mTPI | mTPI-2 | 3+3 | BOIN | BLRM | CRM | i3+3 | mTPI | mTPI-2 | 3+3 | BOIN | BLRM | CRM | i3+3 | mTPI | mTPI-2 | 3+3 | BOIN | BLRM | CRM |
| 1 | 0.01 | 0.319 | 0.309 | 0.333 | 0.604 | 0.164 | 0.217 | 0.184 | 7.737 | 6.573 | 8.259 | 4.839 | 8.043 | 7.257 | 7.212 | 0.065 | 0.069 | 0.087 | 0.047 | 0.075 | 0.071 | 0.071 |
| 2 | 0.3 | 0.636 | 0.656 | 0.627 | 0.358 | 0.753 | 0.747 | 0.73 | 16.647 | 18.315 | 16.524 | 5.253 | 16.572 | 17.859 | 15.114 | 4.919 | 5.448 | 4.916 | 1.632 | 4.918 | 5.336 | 4.512 |
| 3 | 0.55 | 0.043 | 0.033 | 0.037 | 0.036 | 0.079 | 0.036 | 0.086 | 5.088 | 4.737 | 4.836 | 2.049 | 4.974 | 4.551 | 6.66 | 2.797 | 2.63 | 2.673 | 1.121 | 2.737 | 2.554 | 3.664 |
| 4 | 0.65 | 0.002 | 0.002 | 0.003 | 0.002 | 0.004 | 0 | 0 | 0.516 | 0.357 | 0.363 | 0.258 | 0.396 | 0.315 | 0.945 | 0.329 | 0.23 | 0.234 | 0.174 | 0.258 | 0.196 | 0.636 |
| 5 | 0.8 | 0 | 0 | 0 | 0 | 0 | 0 | 0 | 0.012 | 0.018 | 0.018 | 0.015 | 0.015 | 0.018 | 0.069 | 0.008 | 0.012 | 0.012 | 0.011 | 0.014 | 0.016 | 0.053 |
| 6 | 0.95 | 0 | 0 | 0 | 0 | 0 | 0 | 0 | 0 | 0 | 0 | 0 | 0 | 0 | 0 | 0 | 0 | 0 | 0 | 0 | 0 | 0 |

| | i3+3 | mTPI | mTPI-2 | 3+3 | BOIN | BLRM | CRM |
|---|---|---|---|---|---|---|---|
| Prob. of Select MTD | 0.66 | 0.656 | 0.627 | 0.358 | 0.753 | 0.747 | 0.73 |
| Prob. of Toxicity | 0.271 | 0.28 | 0.264 | 0.24 | 0.267 | 0.272 | 0.298 |
| Prob. of Select Dose-over-MTD | 0.045 | 0.035 | 0.04 | 0.038 | 0.083 | 0.036 | 0.086 |
| Prob. of No Selection | 0 | 0 | 0 | 0 | 0 | 0 | 0 |

## Scenario 4

Target Toxicity Prob. = 0.3

| Dose Level | True Tox Prob. | Selection Prob. | | | | | | | # of Patients Treated | | | | | | | # of Toxicities | | | | | | |
|---|---|---|---|---|---|---|---|---|---|---|---|---|---|---|---|---|---|---|---|---|---|---|
| | | i3+3 | mTPI | mTPI-2 | 3+3 | BOIN | BLRM | CRM | i3+3 | mTPI | mTPI-2 | 3+3 | BOIN | BLRM | CRM | i3+3 | mTPI | mTPI-2 | 3+3 | BOIN | BLRM | CRM |
| 1 | 0.04 | 0.002 | 0.001 | 0.001 | 0.031 | 0 | 0 | 0.001 | 3.492 | 3.558 | 3.522 | 3.414 | 3.438 | 3.303 | 3.474 | 0.139 | 0.152 | 0.149 | 0.133 | 0.128 | 0.12 | 0.148 |
| 2 | 0.06 | 0.003 | 0.002 | 0.003 | 0.064 | 0.003 | 0.002 | 0.002 | 3.858 | 3.927 | 3.807 | 3.591 | 3.864 | 3.642 | 3.162 | 0.249 | 0.242 | 0.242 | 0.211 | 0.238 | 0.173 | 0.197 |
| 3 | 0.08 | 0.011 | 0.022 | 0.011 | 0.089 | 0.011 | 0.028 | 0.01 | 4.185 | 4.431 | 4.161 | 3.687 | 4.17 | 5.178 | 3.6 | 0.356 | 0.37 | 0.341 | 0.305 | 0.343 | 0.414 | 0.303 |
| 4 | 0.1 | 0.323 | 0.304 | 0.323 | 0.436 | 0.197 | 0.337 | 0.214 | 6.348 | 5.739 | 6.279 | 4.422 | 6.393 | 8.208 | 6.09 | 0.615 | 0.57 | 0.622 | 0.456 | 0.61 | 0.777 | 0.607 |
| 5 | 0.3 | 0.616 | 0.626 | 0.614 | 0.324 | 0.696 | 0.538 | 0.697 | 9.123 | 9.372 | 9.078 | 4.224 | 9.081 | 7.158 | 9.87 | 2.734 | 2.75 | 2.655 | 1.228 | 2.717 | 2.161 | 2.899 |
| 6 | 0.6 | 0.045 | 0.045 | 0.048 | 0.038 | 0.093 | 0.09 | 0.076 | 2.994 | 2.973 | 3.153 | 1.578 | 3.054 | 2.379 | 3.804 | 1.797 | 1.773 | 1.886 | 0.967 | 1.807 | 1.437 | 2.293 |

| | i3+3 | mTPI | mTPI-2 | 3+3 | BOIN | BLRM | CRM |
|---|---|---|---|---|---|---|---|
| Prob. of Select MTD | 0.616 | 0.626 | 0.614 | 0.324 | 0.696 | 0.538 | 0.697 |
| Prob. of Toxicity | 0.196 | 0.195 | 0.197 | 0.158 | 0.195 | 0.17 | 0.215 |
| Prob. of Select Dose-over-MTD | 0.045 | 0.045 | 0.048 | 0.038 | 0.093 | 0.09 | 0.076 |
| Prob. of No Selection | 0 | 0 | 0 | 0.018 | 0 | 0.005 | 0 |

## Scenario 5

Target Toxicity Prob. = 0.3

| Dose Level | True Tox Prob. | Selection Prob. | | | | | | | # of Patients Treated | | | | | | | # of Toxicities | | | | | | |
|---|---|---|---|---|---|---|---|---|---|---|---|---|---|---|---|---|---|---|---|---|---|---|
| | | i3+3 | mTPI | mTPI-2 | 3+3 | BOIN | BLRM | CRM | i3+3 | mTPI | mTPI-2 | 3+3 | BOIN | BLRM | CRM | i3+3 | mTPI | mTPI-2 | 3+3 | BOIN | BLRM | CRM |
| 1 | 0.05 | 0.991 | 0.99 | 0.982 | 0.929 | 0.886 | 0.93 | 0.919 | 21.477 | 20.826 | 21.345 | 5.88 | 20.922 | 20.937 | 20.436 | 1.075 | 1.102 | 1.114 | 0.287 | 1.005 | 1.011 | 1.052 |
| 2 | 0.6 | 0.009 | 0.009 | 0.017 | 0.042 | 0.114 | 0.049 | 0.08 | 8.121 | 8.739 | 8.226 | 3.924 | 8.682 | 7.842 | 8.199 | 4.884 | 5.258 | 4.943 | 2.4 | 5.155 | 4.73 | 4.915 |
| 3 | 0.8 | 0 | 0 | 0 | 0 | 0 | 0 | 0 | 0.402 | 0.411 | 0.405 | 0.255 | 0.396 | 0.72 | 1.317 | 0.319 | 0.327 | 0.321 | 0.197 | 0.315 | 0.569 | 1.057 |
| 4 | 0.9 | 0 | 0 | 0 | 0 | 0 | 0 | 0 | 0 | 0 | 0.006 | 0.006 | 0.003 | 0.009 | 0.03 | 0 | 0 | 0.006 | 0.006 | 0.003 | 0.009 | 0.029 |
| 5 | 0.95 | 0 | 0 | 0 | 0 | 0 | 0 | 0 | 0 | 0 | 0 | 0 | 0 | 0 | 0 | 0 | 0 | 0 | 0 | 0 | 0 | 0 |
| 6 | 0.99 | 0 | 0 | 0 | 0 | 0 | 0 | 0 | 0 | 0 | 0 | 0 | 0 | 0 | 0 | 0 | 0 | 0 | 0 | 0 | 0 | 0 |

| | i3+3 | mTPI | mTPI-2 | 3+3 | BOIN | BLRM | CRM |
|---|---|---|---|---|---|---|---|
| Prob. of Select MTD | 0.991 | 0.99 | 0.982 | 0.929 | 0.886 | 0.93 | 0.919 |
| Prob. of Toxicity | 0.209 | 0.223 | 0.213 | 0.287 | 0.216 | 0.214 | 0.235 |
| Prob. of Select Dose-over-MTD | 0.009 | 0.009 | 0.017 | 0.042 | 0.114 | 0.049 | 0.08 |
| Prob. of No Selection | 0 | 0.001 | 0.001 | 0.029 | 0 | 0.021 | 0.001 |

## Scenario 6

| Target Toxicity Prob. = 0.3 | | Selection Prob. | | | | | | # of Patients Treated | | | | | | | # of Toxicities | | | | | |
|---|---|---|---|---|---|---|---|---|---|---|---|---|---|---|---|---|---|---|---|---|
| Dose Level | True Tox Prob. | i3+3 | mTPI | mTPI-2 | 3+3 | BOIN | BLRM | CRM | i3+3 | mTPI | mTPI-2 | 3+3 | BOIN | BLRM | CRM | i3+3 | mTPI | mTPI-2 | 3+3 | BOIN | BLRM | CRM |
| 1 | 0.01 | 0.001 | 0.001 | 0.001 | 0.025 | 0 | 0.012 | 0.001 | 3.144 | 3.147 | 3.147 | 3.153 | 3.102 | 3.132 | 3.129 | 0.032 | 0.034 | 0.034 | 0.03 | 0.023 | 0.036 | 0.034 |
| 2 | 0.05 | 0.003 | 0.003 | 0.006 | 0.126 | 0.005 | 0.003 | 0.006 | 3.732 | 3.75 | 3.732 | 3.765 | 3.819 | 3.789 | 3.258 | 0.201 | 0.177 | 0.178 | 0.191 | 0.196 | 0.17 | 0.151 |
| 3 | 0.1 | 0.963 | 0.952 | 0.96 | 0.814 | 0.826 | 0.845 | 0.878 | 15.45 | 14.934 | 15.357 | 5.685 | 15.414 | 16.728 | 14.763 | 1.488 | 1.53 | 1.566 | 0.602 | 1.49 | 1.616 | 1.517 |
| 4 | 0.6 | 0.033 | 0.042 | 0.032 | 0.032 | 0.169 | 0.14 | 0.114 | 7.254 | 7.719 | 7.35 | 3.495 | 7.335 | 6.024 | 7.737 | 4.382 | 4.605 | 4.403 | 2.121 | 4.388 | 3.624 | 4.627 |
| 5 | 0.7 | 0 | 0.002 | 0.001 | 0.001 | 0 | 0 | 0.001 | 0.405 | 0.435 | 0.399 | 0.237 | 0.324 | 0.318 | 1.062 | 0.29 | 0.299 | 0.277 | 0.164 | 0.238 | 0.228 | 0.769 |
| 6 | 0.9 | 0 | 0 | 0 | 0 | 0 | 0 | 0 | 0.015 | 0.015 | 0.015 | 0.006 | 0.006 | 0.009 | 0.051 | 0.015 | 0.012 | 0.012 | 0.005 | 0.005 | 0.009 | 0.04 |
| | | i3+3 | mTPI | mTPI-2 | 3+3 | BOIN | BLRM | CRM | | | | | | | | | | | | | | |
| Prob. of Select MTD | | 0.963 | 0.952 | 0.96 | 0.814 | 0.826 | 0.845 | 0.878 | | | | | | | | | | | | | | |
| Prob. of Toxicity | | 0.214 | 0.222 | 0.216 | 0.191 | 0.211 | 0.189 | 0.238 | | | | | | | | | | | | | | |
| Prob. of Select Dose-over-MTD | | 0.033 | 0.044 | 0.033 | 0.033 | 0.169 | 0.14 | 0.115 | | | | | | | | | | | | | | |
| Prob. of No Selection | | 0 | 0 | 0 | 0 | 0.002 | 0 | 0 | | | | | | | | | | | | | | |

## Scenario 7

| Target Toxicity Prob. = 0.3 | | Selection Prob. | | | | | | # of Patients Treated | | | | | | | # of Toxicities | | | | | |
|---|---|---|---|---|---|---|---|---|---|---|---|---|---|---|---|---|---|---|---|---|
| Dose Level | True Tox Prob. | i3+3 | mTPI | mTPI-2 | 3+3 | BOIN | BLRM | CRM | i3+3 | mTPI | mTPI-2 | 3+3 | BOIN | BLRM | CRM | i3+3 | mTPI | mTPI-2 | 3+3 | BOIN | BLRM | CRM |
| 1 | 0.01 | 0 | 0.001 | 0.001 | 0.011 | 0 | 0 | 0.001 | 3.132 | 3.132 | 3.126 | 3.114 | 3.075 | 3.102 | 3.129 | 0.04 | 0.034 | 0.034 | 0.033 | 0.023 | 0.036 | 0.034 |
| 2 | 0.03 | 0.001 | 0 | 0 | 0.051 | 0.001 | 0.002 | 0.001 | 3.399 | 3.327 | 3.321 | 3.408 | 3.387 | 3.399 | 3.027 | 0.11 | 0.084 | 0.083 | 0.111 | 0.1 | 0.099 | 0.07 |
| 3 | 0.07 | 0.007 | 0.015 | 0.009 | 0.092 | 0.005 | 0.014 | 0.002 | 4.032 | 4.248 | 4.041 | 3.828 | 4.005 | 4.83 | 3.231 | 0.295 | 0.317 | 0.303 | 0.292 | 0.289 | 0.321 | 0.259 |
| 4 | 0.1 | 0.061 | 0.079 | 0.058 | 0.187 | 0.031 | 0.187 | 0.034 | 4.692 | 4.728 | 4.671 | 3.942 | 4.704 | 6.966 | 4.08 | 0.454 | 0.455 | 0.45 | 0.399 | 0.465 | 0.639 | 0.404 |
| 5 | 0.15 | 0.913 | 0.891 | 0.92 | 0.648 | 0.922 | 0.712 | 0.92 | 10.554 | 10.761 | 10.839 | 4.884 | 10.707 | 8.373 | 12.159 | 1.602 | 1.6 | 1.612 | 0.748 | 1.583 | 1.237 | 1.779 |
| 6 | 0.75 | 0.018 | 0.014 | 0.012 | 0.007 | 0.041 | 0.085 | 0.042 | 4.191 | 3.804 | 4.002 | 2.37 | 4.122 | 3.33 | 4.374 | 3.068 | 2.854 | 3.001 | 1.788 | 3.089 | 2.487 | 3.291 |
| | | i3+3 | mTPI | mTPI-2 | 3+3 | BOIN | BLRM | CRM | | | | | | | | | | | | | | |
| Prob. of Select MTD | | 0.913 | 0.891 | 0.92 | 0.648 | 0.922 | 0.712 | 0.92 | | | | | | | | | | | | | | |
| Prob. of Toxicity | | 0.186 | 0.178 | 0.183 | 0.156 | 0.185 | 0.161 | 0.195 | | | | | | | | | | | | | | |
| Prob. of Select Dose-over-MTD | | 0.018 | 0.014 | 0.012 | 0.007 | 0.041 | 0.085 | 0.042 | | | | | | | | | | | | | | |
| Prob. of No Selection | | 0 | 0 | 0 | 0.004 | 0 | 0 | 0 | | | | | | | | | | | | | | |

## Scenario 8

| Target Toxicity Prob. = 0.3 | | Selection Prob. | | | | | | # of Patients Treated | | | | | | | # of Toxicities | | | | | |
|---|---|---|---|---|---|---|---|---|---|---|---|---|---|---|---|---|---|---|---|---|
| Dose Level | True Tox Prob. | i3+3 | mTPI | mTPI-2 | 3+3 | BOIN | BLRM | CRM | i3+3 | mTPI | mTPI-2 | 3+3 | BOIN | BLRM | CRM | i3+3 | mTPI | mTPI-2 | 3+3 | BOIN | BLRM | CRM |
| 1 | 0.29 | 0.351 | 0.38 | 0.346 | 0.256 | 0.356 | 0.22 | 0.344 | 14.202 | 14.883 | 14.487 | 4.968 | 14.16 | 9.213 | 13.362 | 4.038 | 4.274 | 4.164 | 1.421 | 4.101 | 2.694 | 3.895 |
| 2 | 0.31 | 0.197 | 0.199 | 0.185 | 0.118 | 0.224 | 0.236 | 0.2 | 7.164 | 7.119 | 7.044 | 2.595 | 6.894 | 7.716 | 6.282 | 2.289 | 2.236 | 2.201 | 0.834 | 2.104 | 2.344 | 1.945 |
| 3 | 0.33 | 0.118 | 0.123 | 0.144 | 0.07 | 0.131 | 0.124 | 0.182 | 3.315 | 3.435 | 3.588 | 1.182 | 3.597 | 3.645 | 4.557 | 1.087 | 1.073 | 1.136 | 0.383 | 1.182 | 1.193 | 1.443 |
| 4 | 0.35 | 0.063 | 0.052 | 0.07 | 0.017 | 0.079 | 0.038 | 0.091 | 1.41 | 1.284 | 1.56 | 0.558 | 1.611 | 1.017 | 2.166 | 0.495 | 0.456 | 0.558 | 0.22 | 0.557 | 0.36 | 0.75 |
| 5 | 0.37 | 0.028 | 0.027 | 0.026 | 0.005 | 0.036 | 0.009 | 0.038 | 0.498 | 0.426 | 0.477 | 0.171 | 0.588 | 0.243 | 0.801 | 0.186 | 0.157 | 0.172 | 0.064 | 0.212 | 0.08 | 0.302 |
| 6 | 0.39 | 0.014 | 0.007 | 0.016 | 0.005 | 0.017 | 0.014 | 0.008 | 0.123 | 0.102 | 0.159 | 0.045 | 0.195 | 0.132 | 0.177 | 0.046 | 0.048 | 0.066 | 0.016 | 0.076 | 0.046 | 0.079 |
| | | i3+3 | mTPI | mTPI-2 | 3+3 | BOIN | BLRM | CRM | | | | | | | | | | | | | | |
| Prob. of Select MTD | | 0.729 | 0.754 | 0.745 | 0.461 | 0.79 | 0.618 | 0.817 | | | | | | | | | | | | | | |
| Prob. of Toxicity | | 0.305 | 0.303 | 0.304 | 0.309 | 0.304 | 0.306 | 0.308 | | | | | | | | | | | | | | |
| Prob. of Select Dose-over-MTD | | 0.042 | 0.034 | 0.042 | 0.01 | 0.053 | 0.023 | 0.046 | | | | | | | | | | | | | | |
| Prob. of No Selection | | 0.229 | 0.212 | 0.213 | 0.529 | 0.157 | 0.359 | 0.137 | | | | | | | | | | | | | | |

## Scenario 9

| Target Toxicity Prob. = 0.3 | | Selection Prob. | | | | | | # of Patients Treated | | | | | | | # of Toxicities | | | | | |
|---|---|---|---|---|---|---|---|---|---|---|---|---|---|---|---|---|---|---|---|---|
| Dose Level | True Tox Prob. | i3+3 | mTPI | mTPI-2 | 3+3 | BOIN | BLRM | CRM | i3+3 | mTPI | mTPI-2 | 3+3 | BOIN | BLRM | CRM | i3+3 | mTPI | mTPI-2 | 3+3 | BOIN | BLRM | CRM |
| 1 | 0.25 | 0.27 | 0.329 | 0.286 | 0.247 | 0.256 | 0.209 | 0.218 | 12.093 | 13.023 | 12.438 | 4.872 | 11.829 | 9.207 | 11.019 | 3.015 | 3.198 | 3.031 | 1.229 | 2.888 | 2.266 | 2.708 |
| 2 | 0.27 | 0.219 | 0.252 | 0.23 | 0.161 | 0.217 | 0.273 | 0.258 | 7.749 | 8.34 | 7.914 | 2.925 | 7.71 | 8.361 | 7.275 | 2.079 | 2.311 | 2.209 | 0.809 | 2.095 | 2.282 | 1.999 |
| 3 | 0.29 | 0.166 | 0.152 | 0.164 | 0.077 | 0.2 | 0.165 | 0.204 | 4.464 | 4.11 | 4.431 | 1.623 | 4.737 | 4.749 | 5.235 | 1.272 | 1.169 | 1.282 | 0.496 | 1.365 | 1.428 | 1.514 |
| 4 | 0.31 | 0.125 | 0.086 | 0.108 | 0.037 | 0.126 | 0.078 | 0.141 | 2.415 | 1.809 | 2.151 | 0.78 | 2.37 | 1.758 | 3.111 | 0.751 | 0.551 | 0.655 | 0.258 | 0.723 | 0.546 | 0.948 |
| 5 | 0.33 | 0.065 | 0.042 | 0.058 | 0.016 | 0.071 | 0.03 | 0.069 | 0.948 | 0.738 | 1.029 | 0.36 | 1.137 | 0.489 | 1.209 | 0.335 | 0.23 | 0.328 | 0.12 | 0.365 | 0.15 | 0.388 |
| 6 | 0.35 | 0.029 | 0.028 | 0.045 | 0.013 | 0.044 | 0.02 | 0.033 | 0.351 | 0.354 | 0.441 | 0.135 | 0.501 | 0.219 | 0.555 | 0.133 | 0.124 | 0.148 | 0.05 | 0.187 | 0.076 | 0.186 |
| | | i3+3 | mTPI | mTPI-2 | 3+3 | BOIN | BLRM | CRM | | | | | | | | | | | | | | |
| Prob. of Select MTD | | 0.874 | 0.889 | 0.891 | 0.551 | 0.914 | 0.775 | 0.923 | | | | | | | | | | | | | | |
| Prob. of Toxicity | | 0.271 | 0.267 | 0.269 | 0.277 | 0.27 | 0.272 | 0.273 | | | | | | | | | | | | | | |
| Prob. of Select Dose-over-MTD | | 0 | 0 | 0 | 0 | 0 | 0 | 0 | | | | | | | | | | | | | | |
| Prob. of No Selection | | 0.126 | 0.111 | 0.109 | 0.449 | 0.086 | 0.225 | 0.077 | | | | | | | | | | | | | | |

## Scenario 10

| Target Toxicity Prob. = 0.3 | | Selection Prob. | | | | | | # of Patients Treated | | | | | | | # of Toxicities | | | | | |
|---|---|---|---|---|---|---|---|---|---|---|---|---|---|---|---|---|---|---|---|---|
| Dose Level | True Tox Prob. | i3+3 | mTPI | mTPI-2 | 3+3 | BOIN | BLRM | CRM | i3+3 | mTPI | mTPI-2 | 3+3 | BOIN | BLRM | CRM | i3+3 | mTPI | mTPI-2 | 3+3 | BOIN | BLRM | CRM |
| 1 | 0.21 | 0.158 | 0.218 | 0.163 | 0.244 | 0.151 | 0.14 | 0.126 | 9.444 | 10.587 | 9.552 | 4.767 | 9.564 | 7.551 | 8.622 | 1.95 | 2.161 | 1.967 | 1.027 | 1.983 | 1.539 | 1.765 |
| 2 | 0.23 | 0.203 | 0.247 | 0.209 | 0.153 | 0.202 | 0.232 | 0.21 | 7.551 | 8.31 | 7.851 | 3.27 | 7.596 | 8.271 | 6.879 | 1.715 | 1.925 | 1.799 | 0.788 | 1.699 | 1.927 | 1.602 |
| 3 | 0.25 | 0.199 | 0.211 | 0.215 | 0.11 | 0.204 | 0.231 | 0.233 | 5.616 | 5.361 | 5.706 | 2.031 | 5.643 | 6.288 | 6.231 | 1.403 | 1.349 | 1.429 | 0.493 | 1.403 | 1.552 | 1.575 |
| 4 | 0.27 | 0.179 | 0.138 | 0.152 | 0.072 | 0.174 | 0.148 | 0.185 | 3.561 | 2.811 | 3.411 | 1.281 | 3.618 | 2.856 | 4.044 | 0.981 | 0.767 | 0.939 | 0.35 | 0.979 | 0.802 | 1.087 |
| 5 | 0.29 | 0.107 | 0.071 | 0.124 | 0.031 | 0.127 | 0.058 | 0.133 | 1.797 | 1.338 | 1.68 | 0.696 | 1.8 | 0.981 | 2.205 | 0.548 | 0.39 | 0.469 | 0.214 | 0.512 | 0.288 | 0.63 |
| 6 | 0.31 | 0.089 | 0.067 | 0.09 | 0.034 | 0.109 | 0.049 | 0.078 | 0.912 | 0.783 | 0.99 | 0.288 | 1.089 | 0.624 | 1.209 | 0.297 | 0.241 | 0.313 | 0.088 | 0.362 | 0.213 | 0.364 |
| | | i3+3 | mTPI | mTPI-2 | 3+3 | BOIN | BLRM | CRM | | | | | | | | | | | | | | |
| Prob. of Select MTD | | 0.574 | 0.487 | 0.581 | 0.247 | 0.614 | 0.486 | 0.629 | | | | | | | | | | | | | | |
| Prob. of Toxicity | | 0.239 | 0.234 | 0.237 | 0.24 | 0.237 | 0.238 | 0.241 | | | | | | | | | | | | | | |
| Prob. of Select Dose-over-MTD | | 0 | 0 | 0 | 0 | 0 | 0 | 0 | | | | | | | | | | | | | | |
| Prob. of No Selection | | 0.065 | 0.048 | 0.047 | 0.356 | 0.033 | 0.142 | 0.035 | | | | | | | | | | | | | | |

### Scenario 11
Target Toxicity Prob. = 0.3

| Dose Level | True Tox Prob. | Selection Prob. | | | | | | | # of Patients Treated | | | | | | | # of Toxicities | | | | | |
|---|---|---|---|---|---|---|---|---|---|---|---|---|---|---|---|---|---|---|---|---|---|
| | | i3+3 | mTPI | mTPI-2 | 3+3 | BOIN | BLRM | CRM | i3+3 | mTPI | mTPI-2 | 3+3 | BOIN | BLRM | CRM | i3+3 | mTPI | mTPI-2 | 3+3 | BOIN | BLRM | CRM |
| 1 | 0.05 | 0.067 | 0.057 | 0.071 | 0.328 | 0.042 | 0.032 | 0.043 | 4.986 | 4.641 | 5.16 | 4.269 | 5.085 | 4.434 | 4.521 | 0.236 | 0.243 | 0.263 | 0.199 | 0.244 | 0.189 | 0.233 |
| 2 | 0.2 | 0.274 | 0.276 | 0.252 | 0.298 | 0.207 | 0.309 | 0.163 | 9.543 | 9.933 | 9.273 | 4.761 | 9.183 | 9.969 | 6.66 | 1.947 | 2.013 | 1.89 | 0.986 | 1.812 | 1.969 | 1.374 |
| 3 | 0.27 | 0.329 | 0.338 | 0.309 | 0.204 | 0.352 | 0.4 | 0.318 | 8.466 | 8.529 | 8.064 | 3.366 | 8.265 | 9.807 | 8.763 | 2.337 | 2.259 | 2.141 | 0.91 | 2.208 | 2.679 | 2.408 |
| 4 | 0.33 | 0.196 | 0.228 | 0.232 | 0.107 | 0.243 | 0.185 | 0.309 | 4.704 | 4.548 | 4.848 | 1.932 | 4.995 | 4.221 | 6.201 | 1.589 | 1.469 | 1.562 | 0.639 | 1.656 | 1.385 | 1.979 |
| 5 | 0.39 | 0.099 | 0.079 | 0.101 | 0.035 | 0.116 | 0.051 | 0.137 | 1.734 | 1.911 | 2.097 | 0.831 | 1.968 | 1.149 | 2.964 | 0.66 | 0.745 | 0.826 | 0.337 | 0.735 | 0.452 | 1.126 |
| 6 | 0.45 | 0.035 | 0.021 | 0.034 | 0.004 | 0.04 | 0.017 | 0.03 | 0.567 | 0.42 | 0.54 | 0.24 | 0.504 | 0.261 | 0.891 | 0.243 | 0.191 | 0.241 | 0.121 | 0.229 | 0.123 | 0.413 |
| | | i3+3 | mTPI | mTPI-2 | 3+3 | BOIN | BLRM | CRM | | | | | | | | | | | | | | |
| Prob. of Select MTD | | 0.525 | 0.566 | 0.541 | 0.311 | 0.595 | 0.585 | 0.627 | | | | | | | | | | | | | | |
| Prob. of Toxicity | | 0.234 | 0.231 | 0.231 | 0.207 | 0.229 | 0.228 | 0.251 | | | | | | | | | | | | | | |
| Prob. of Select Dose-over-MTD | | 0.134 | 0.1 | 0.135 | 0.039 | 0.156 | 0.068 | 0.167 | | | | | | | | | | | | | | |
| Prob. of No Selection | | 0 | 0.001 | 0.001 | 0.024 | 0 | 0.006 | 0 | | | | | | | | | | | | | | |

### Scenario 12
Target Toxicity Prob. = 0.3

| Dose Level | True Tox Prob. | Selection Prob. | | | | | | | # of Patients Treated | | | | | | | # of Toxicities | | | | | |
|---|---|---|---|---|---|---|---|---|---|---|---|---|---|---|---|---|---|---|---|---|---|
| | | i3+3 | mTPI | mTPI-2 | 3+3 | BOIN | BLRM | CRM | i3+3 | mTPI | mTPI-2 | 3+3 | BOIN | BLRM | CRM | i3+3 | mTPI | mTPI-2 | 3+3 | BOIN | BLRM | CRM |
| 1 | 0.05 | 0.007 | 0.004 | 0.005 | 0.087 | 0.002 | 0.001 | 0.004 | 3.864 | 3.834 | 3.81 | 3.615 | 3.711 | 3.477 | 3.693 | 0.21 | 0.201 | 0.2 | 0.188 | 0.178 | 0.147 | 0.195 |
| 2 | 0.1 | 0.071 | 0.068 | 0.072 | 0.29 | 0.04 | 0.086 | 0.034 | 5.517 | 5.55 | 5.562 | 4.392 | 5.604 | 5.802 | 4.053 | 0.555 | 0.565 | 0.571 | 0.43 | 0.552 | 0.54 | 0.422 |
| 3 | 0.2 | 0.337 | 0.382 | 0.354 | 0.308 | 0.297 | 0.408 | 0.237 | 8.346 | 9.312 | 8.793 | 4.428 | 8.583 | 10.845 | 7.368 | 1.625 | 1.893 | 1.743 | 0.912 | 1.664 | 2.22 | 1.483 |
| 4 | 0.3 | 0.362 | 0.373 | 0.357 | 0.218 | 0.406 | 0.383 | 0.455 | 7.764 | 7.425 | 7.536 | 3.153 | 7.752 | 7.119 | 8.883 | 2.341 | 2.212 | 2.269 | 0.93 | 2.321 | 2.119 | 2.61 |
| 5 | 0.4 | 0.146 | 0.144 | 0.17 | 0.053 | 0.173 | 0.098 | 0.232 | 3.42 | 3.132 | 3.384 | 1.53 | 3.333 | 2.133 | 4.68 | 1.376 | 1.24 | 1.347 | 0.629 | 1.317 | 0.89 | 1.869 |
| 6 | 0.5 | 0.077 | 0.029 | 0.042 | 0.012 | 0.082 | 0.018 | 0.038 | 1.089 | 0.747 | 0.915 | 0.366 | 1.017 | 0.465 | 1.323 | 0.421 | 0.356 | 0.452 | 0.199 | 0.393 | 0.235 | 0.665 |
| | | i3+3 | mTPI | mTPI-2 | 3+3 | BOIN | BLRM | CRM | | | | | | | | | | | | | | |
| Prob. of Select MTD | | 0.362 | 0.373 | 0.357 | 0.218 | 0.406 | 0.383 | 0.455 | | | | | | | | | | | | | | |
| Prob. of Toxicity | | 0.218 | 0.216 | 0.219 | 0.188 | 0.214 | 0.206 | 0.241 | | | | | | | | | | | | | | |
| Prob. of Select Dose-over-MTD | | 0.223 | 0.173 | 0.212 | 0.065 | 0.255 | 0.116 | 0.27 | | | | | | | | | | | | | | |
| Prob. of No Selection | | 0 | 0 | 0 | 0.032 | 0 | 0.006 | 0 | | | | | | | | | | | | | | |

### Scenario 13
Target Toxicity Prob. = 0.3

| Dose Level | True Tox Prob. | Selection Prob. | | | | | | | # of Patients Treated | | | | | | | # of Toxicities | | | | | |
|---|---|---|---|---|---|---|---|---|---|---|---|---|---|---|---|---|---|---|---|---|---|
| | | i3+3 | mTPI | mTPI-2 | 3+3 | BOIN | BLRM | CRM | i3+3 | mTPI | mTPI-2 | 3+3 | BOIN | BLRM | CRM | i3+3 | mTPI | mTPI-2 | 3+3 | BOIN | BLRM | CRM |
| 1 | 0.3 | 0.447 | 0.427 | 0.408 | 0.288 | 0.447 | 0.297 | 0.421 | 16.26 | 15.867 | 15.981 | 5.055 | 15.576 | 10.725 | 14.733 | 4.939 | 4.754 | 4.779 | 1.488 | 4.592 | 3.284 | 4.41 |
| 2 | 0.35 | 0.168 | 0.225 | 0.214 | 0.105 | 0.247 | 0.199 | 0.258 | 6.717 | 7.653 | 7.263 | 2.469 | 7.443 | 6.855 | 6.906 | 2.406 | 2.655 | 2.532 | 0.894 | 2.604 | 2.454 | 2.401 |
| 3 | 0.4 | 0.091 | 0.069 | 0.103 | 0.042 | 0.099 | 0.068 | 0.119 | 2.619 | 2.592 | 2.898 | 0.918 | 2.73 | 2.496 | 3.9 | 1.01 | 1.031 | 1.129 | 0.37 | 1.107 | 0.999 | 1.53 |
| 4 | 0.45 | 0.023 | 0.02 | 0.021 | 0.008 | 0.028 | 0.009 | 0.034 | 0.768 | 0.657 | 0.768 | 0.291 | 0.828 | 0.492 | 1.125 | 0.344 | 0.295 | 0.369 | 0.139 | 0.362 | 0.218 | 0.533 |
| 5 | 0.5 | 0.006 | 0.001 | 0.002 | 0.001 | 0.01 | 0.003 | 0.003 | 0.168 | 0.117 | 0.12 | 0.051 | 0.189 | 0.096 | 0.267 | 0.077 | 0.068 | 0.062 | 0.026 | 0.086 | 0.05 | 0.133 |
| 6 | 0.55 | 0.002 | 0 | 0 | 0 | 0 | 0 | 0 | 0.024 | 0 | 0.003 | 0.003 | 0.012 | 0.009 | 0.015 | 0.011 | 0 | 0.002 | 0.002 | 0.009 | 0.005 | 0.011 |
| | | i3+3 | mTPI | mTPI-2 | 3+3 | BOIN | BLRM | CRM | | | | | | | | | | | | | | |
| Prob. of Select MTD | | 0.615 | 0.652 | 0.622 | 0.393 | 0.694 | 0.496 | 0.679 | | | | | | | | | | | | | | |
| Prob. of Toxicity | | 0.331 | 0.327 | 0.328 | 0.332 | 0.327 | 0.339 | 0.335 | | | | | | | | | | | | | | |
| Prob. of Select Dose-over-MTD | | 0.122 | 0.09 | 0.126 | 0.051 | 0.137 | 0.08 | 0.156 | | | | | | | | | | | | | | |
| Prob. of No Selection | | 0.263 | 0.258 | 0.252 | 0.556 | 0.169 | 0.424 | 0.165 | | | | | | | | | | | | | | |

### Scenario 14
Target Toxicity Prob. = 0.3

| Dose Level | True Tox Prob. | Selection Prob. | | | | | | | # of Patients Treated | | | | | | | # of Toxicities | | | | | |
|---|---|---|---|---|---|---|---|---|---|---|---|---|---|---|---|---|---|---|---|---|---|
| | | i3+3 | mTPI | mTPI-2 | 3+3 | BOIN | BLRM | CRM | i3+3 | mTPI | mTPI-2 | 3+3 | BOIN | BLRM | CRM | i3+3 | mTPI | mTPI-2 | 3+3 | BOIN | BLRM | CRM |
| 1 | 0.15 | 0.057 | 0.082 | 0.061 | 0.212 | 0.054 | 0.053 | 0.041 | 6.633 | 7.461 | 6.801 | 4.494 | 6.573 | 5.517 | 6.18 | 1.017 | 1.115 | 1.026 | 0.676 | 0.966 | 0.774 | 0.931 |
| 2 | 0.18 | 0.15 | 0.171 | 0.153 | 0.194 | 0.128 | 0.164 | 0.128 | 7.005 | 7.074 | 7.029 | 3.741 | 7.122 | 7.851 | 5.841 | 1.258 | 1.205 | 1.225 | 0.689 | 1.259 | 1.411 | 1.022 |
| 3 | 0.21 | 0.202 | 0.24 | 0.207 | 0.151 | 0.208 | 0.266 | 0.217 | 6.387 | 6.774 | 6.36 | 2.862 | 6.333 | 7.686 | 6.417 | 1.353 | 1.469 | 1.357 | 0.606 | 1.321 | 1.59 | 1.363 |
| 4 | 0.24 | 0.198 | 0.227 | 0.21 | 0.111 | 0.221 | 0.232 | 0.247 | 4.584 | 4.62 | 4.833 | 1.938 | 4.923 | 4.377 | 5.484 | 1.061 | 1.105 | 1.168 | 0.466 | 1.169 | 1.034 | 1.323 |
| 5 | 0.27 | 0.192 | 0.137 | 0.18 | 0.069 | 0.183 | 0.118 | 0.204 | 2.997 | 2.334 | 2.934 | 1.251 | 2.952 | 1.962 | 3.657 | 0.767 | 0.59 | 0.771 | 0.338 | 0.789 | 0.532 | 0.981 |
| 6 | 0.3 | 0.181 | 0.134 | 0.18 | 0.06 | 0.198 | 0.114 | 0.155 | 1.932 | 1.542 | 1.848 | 0.594 | 1.902 | 1.239 | 2.226 | 0.592 | 0.445 | 0.522 | 0.193 | 0.553 | 0.353 | 0.642 |
| | | i3+3 | mTPI | mTPI-2 | 3+3 | BOIN | BLRM | CRM | | | | | | | | | | | | | | |
| Prob. of Select MTD | | 0.373 | 0.271 | 0.36 | 0.129 | 0.381 | 0.232 | 0.359 | | | | | | | | | | | | | | |
| Prob. of Toxicity | | 0.205 | 0.199 | 0.204 | 0.199 | 0.203 | 0.119 | 0.21 | | | | | | | | | | | | | | |
| Prob. of Select Dose-over-MTD | | 0 | 0 | 0 | 0 | 0 | 0 | 0 | | | | | | | | | | | | | | |
| Prob. of No Selection | | 0.02 | 0.009 | 0.009 | 0.203 | 0.008 | 0.053 | 0.008 | | | | | | | | | | | | | | |

# Appendix D
# Comparison of decision tables for different designs

We provide a quick comparison of decision tables among the i3+3, mTPI, and mTPI-2 designs, since all three designs require the same input of the target toxicity probability $p_T$ and the EI. Figure A1 provides a side-by-side comparison based on $p_T = 0.3$ and EI [0.25, 0.35] for up to 9 patients. The dose assignment decisions for i3+3 seems to be more reasonable than mTPI or mTPI-2. For example, when 1 out of 2 patients experiences DLT, the decision is "S" under i3+3 and mTPI, compared with the decision "D" under mTPI-2. Here, one could argue that when only 2 patients have been treated at a dose, 1 DLT does not provide sufficient information to warrant a de-escalation decision. Investigators might want to wait for further information to make a dose change decision. In this case, i3+3 agrees with mTPI, not mTPI-2. As a contrasting example, when 3 out of 6 patients experiences DLTs, the decision is "D" under i3+3 and mTPI-2, compared with the decision "S" under mTPI. Again, one could argue that "S" is more sensible according to our earlier discussion. Therefore, i3+3 here agrees with mTPI-2, not mTPI. In general, we can see that i3+3 seems to take more sound and ethically acceptable decisions, often complementing the mTPI and mTPI-2 designs.

**Figure A1:** Comparison of the decision tables for i3+3, mTPI and mTPI-2. For each "Number of Patients" (column), there are three subcolumns listing the decisions of i3+3, mTPI, and mTPI-2 side-by-side corresponding to the "Number of DLTs" (row). Here, the target toxicity probability $p_T = 0.3$ and EI (0.25, 0.35).

Other than the slightly different decision tables, all three designs use the same statistical models and inference to select the MTD. As the mTPI and mTPI-2 designs have been well established in the literature with desirable performance (Ji et al., 2010; Guo et al., 2017; Ji and Yang, 2018; Zhou et al, 2018), we decide not to conduct more and larger simulations. The i3+3 design is expected to perform at a similar level as the mTPI and mTPI-2 designs due to its similar but slightly improved decision table, which puts i3+3 in par with other major model-based designs.

While the decisions of the i3+3 design can be calculated and tabulated prior to the trial onset, the decisions of the model-based design such as CRM and BLRM require intensive computation and simulation to investigate. In particular, for each row and column in Figure A2, while the decision of i3+3 is fixed, that of CRM and BLRM is random and takes "D", "S", or "E" following a probability distribution. See Ji and Yang (2018) for a detailed discussion. To investigate the behavior of the CRM and BLRM designs, we summarize the empirical distribution of the random decisions by computing the frequencies of "D", "S", "E" for each column and row value given the observed toxicity data corresponding to the row and column numbers. The frequencies of the three decisions are plotted in terms of horizontal color bars shown in Figure A2, with the bar length proportional to the frequency.

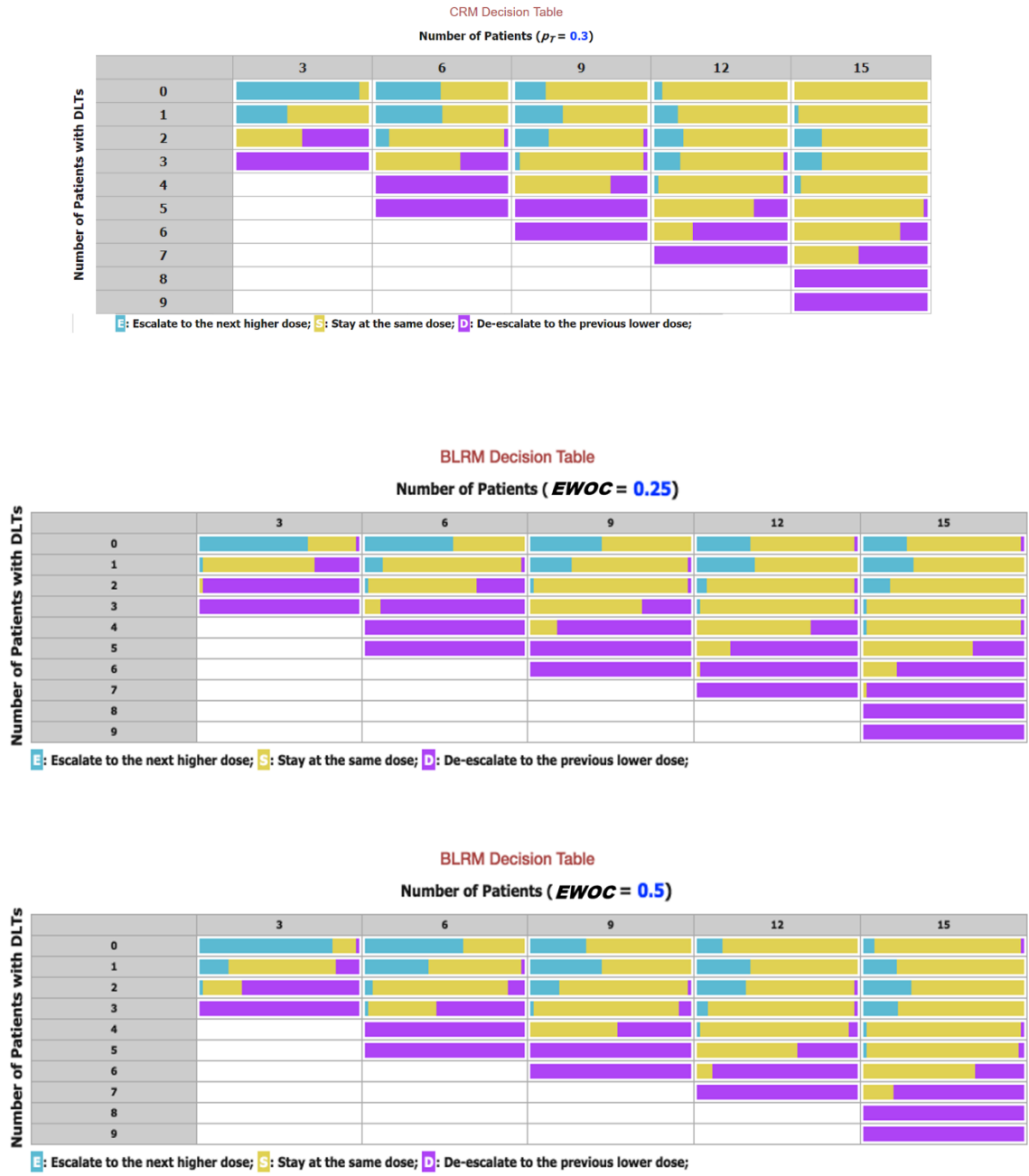

**Figure A2**: Decision tables under the CRM and BLRM design for $p_T = 0.3$. Each column represents $n$ number of patients treated at the current dose and each row represents $x$ number of patients with DLTs. For each entry, the decisions E, S, and D are colored blue, yellow, and purple, respectively. The length of the colored segments within each bar are proportional to the three proportions of the three decisions taken by CRM and BLRM for a given $(x; n)$ data point from a simulation study using the 14 scenarios in Appendix B and 10,000 simulated trials per scenarios.

It can be seen from the table that more than half of the times the CRM design would stay, and about one fifth of the times the BLRM (with EWOC=0.5) design would de-escalate when 1 out of 3 patients experienced DLT at a given dose. When 3 out of 6 patients experienced DLT, about half of the times CRM and BLRM (with EWOC=0.5) would de-escalate but the other half of the times it would stay. These decisions may not pass safety review committees in real-world clinical trialsn real-world clinical trials.